\documentclass{ethpaper}
\usepackage{graphicx}
\newcommand{\be}{\begin{equation}}
\newcommand{\ee}{\end{equation}}
\newcommand{\expfor}[2]{$#1\!\times\! 10^{#2}$}
\newcommand{\muind}{\mu_{\rm IND}(420)}
\newcommand{\muindl}{\mu_{\rm IND}(\lambda)}
\newcommand{\muindhosp}{\mu_{\rm IND}^{\gamma_{\rm std}}(420)}
\newcommand{\muindt}{\mu_{\rm IND}(420,t_{\rm rec})}
\newcommand{\iadr}{\dot D_{\rm ind}}
\newcommand{\tauyks}{17.2}
\newcommand{\taukaks}{650}

\usepackage{amssymb}
\begin{document}
\begin{titlepage}

\ethnote{}
\title{High-energy proton induced damage in PbWO$_4$ calorimeter crystals}
\begin{Authlist}
M.~Huhtinen
\Instfoot{cern}{CERN, CH-1211 Geneva, Switzerland}
P.~Lecomte, D.~Luckey, F.~Nessi-Tedaldi, F.~Pauss
\Instfoot{eth}{Swiss Federal Institute of Technology (ETH),
CH-8093 Z\"urich, Switzerland}
\end{Authlist}
\maketitle

\begin{abstract}
Eight production quality PbWO$_4$ crystals of CMS have been irradiated
in a 20\,GeV/c proton beam up to fluences of
\expfor{5.4}{13}\,cm$^{-2}$. The damage recovery in these crystals has
been followed for over a year. Comparative irradiations with $^{60}$Co
photons have been performed on seven other crystals using a dose rate
of 1\,kGy/h.  In proton irradiated crystals the light transmission
band-edge shifts and the induced absorption length
$\muindl\propto\lambda^{-4}$. In $\gamma$-irradiated crystals the
band-edge does not shift but formation of absorption bands is seen
clearly.  The absorption length induced by $\gamma$-radiation in
crystals verified to have excellent radiation hardness, saturates at a
level below 0.5\,m$^{-1}$. In the case of protons, we observe no
correlation with the pre-characterised radiation hardness of the
crystals and the induced absorption increases linearly with
fluence. After a fluence of \expfor{5}{13}\,cm$^{-2}$, an induced
absorption length of $\sim$15\,m$^{-1}$ is seen with no sign of
saturation. These observations provide strong evidence that
high-energy protons create damage that cannot be reproduced with
$\gamma$-irradiation.  However, these hadronic effects manifest
themselves only at integral fluences well beyond $10^{12}$\,cm$^{-2}$
and most likely would escape undetected at lower fluences.  A large
fraction of the damage, both in proton- and $\gamma$-irradiated
crystals, is either stable or recovers very slowly.
\end{abstract}

\vspace{7cm}
\conference{Submitted to Elsevier Science}

\end{titlepage}

\section{Introduction}

Lead Tungstate (PbWO\,$_4$) crystals will be used by several
high-energy physics experiments \cite{etdr,alice,btev} because they
provide a compact homogeneous calorimeter with fast scintillation.  In
view of their use in the CMS experiment, where the calorimeters will
face the harsh LHC conditions, the radiation hardness of PbWO\,$_4$
crystals has been extensively studied over the last
decade\cite{scint97,ieee47:1741,nim490:30,paul,nim414:149,nim433:630}. These
studies have mostly been done at $\gamma$-irradiation facilities,
complemented with some neutron tests at reactors\,\cite{tn95:126}. At
LHC, however, calorimeters will also be exposed to a large fluence of
high-energy hadrons, resulting from up to \expfor{8}{8}\,s$^{-1}$
minimum bias $pp$ interactions at $\sqrt{s}$=14\,TeV.

Although the radiation damage caused by minimum ionising particles in
PbWO\,$_4$ has been thoroughly studied by the $\gamma$-irradiations,
this cannot be considered sufficient for LHC applications. In
particular, the fundamental difference between few MeV photons and
energetic hadrons -- the effects of nuclear interactions in the
crystal -- has not been investigated in detail so far. Some pion
irradiations have been performed\,\cite{note98:069,nim530:286} but no
systematic study has been extended up to the full integrated hadron
fluences expected at the LHC, although the neccessity of this has been
advocated long ago\,\cite{note98:055}.

\begin{figure}[h]
\begin{center}
\includegraphics*[height=10cm]{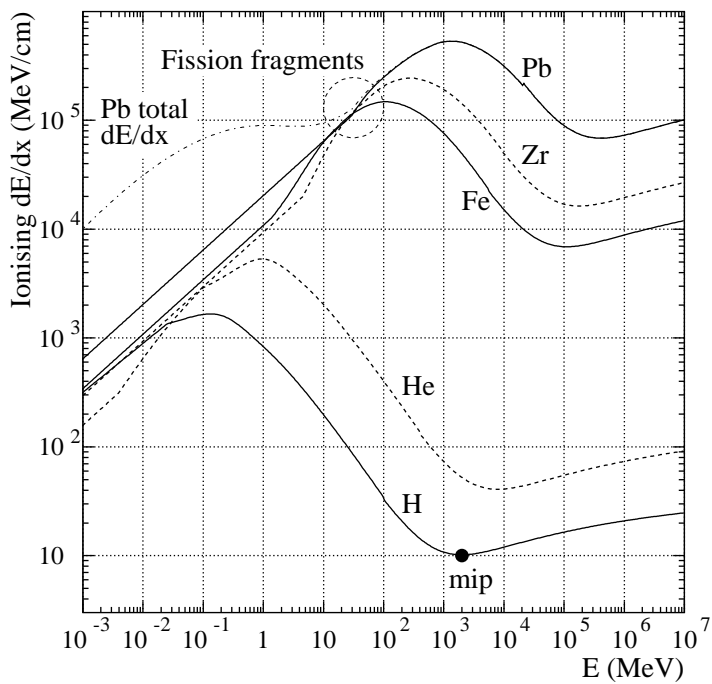}
\end{center}
\caption{Simulated ionising energy loss of different fragments in
  PbWO$_4$. Non-ionising energy loss is not included, but even for
  heavy fragments it is negligible above $\sim$10\,MeV.  This is
  quantified for lead fragments, where the total $dE/dx$ is indicated
  by the dot-dashed line.  The dashed circle shows the typical
  $dE/dx$-values of fission fragments at the beginning of their
  track.}
\label{fig1}
\end{figure}

Inelastic nuclear interactions break up the target nucleus and thus
create impurities and distortions in the crystal lattice. These
effects, however, are expected to be negligible at LHC fluences, since
the natural impurity concentration of the crystals is relatively high.
Another unique feature of hadronic interactions is the very dense
ionisation of the created heavy nuclear fragments, which can have a
range of up to 10\,$\mu$m. Along their path they displace a large
number of lattice atoms, but also ionise much more densely than a
minimum ionising particle.  In PbWO$_4$ the fragments with highest
ionising $dE/dx$ are known to come from fission of lead and tungsten
where the cross section for such reactions exceeds
100\,mb\,\cite{np686:481}.  Typical fission fragments, like Fe and Zr
in Fig\,\ref{fig1}, have energies up to 100\,MeV. It can be seen from
Fig.\,\ref{fig1} that the $dE/dx$ for such fragments is four orders of
magnitude larger than for minimum-ionising protons -- indicated by the
solid dot in Fig\,\ref{fig1}.  Since the thresholds to induce fission
are several hundred MeV, tests with reactor neutrons or low-energy
protons are not suitable to probe this regime.

An irradiation test of PbWO\,$_4$ with hadrons is more complicated
than a standard $\gamma$-irradiation.  The production of nuclear
fragments causes the crystals to become radioactive, up to a level
which does not allow handling them for several weeks after exposure.
Any front-end electronics or optical system, like photomultipliers,
laser or fibre is also susceptible to damage in the hadron beam and
would make the determination of an effect in the crystal itself more
complicated. Therefore our aim was to make the simplest possible test
we could imagine. We irradiated bare crystals and measured the light
transmission after irradiation.  Our study was targeted at stable or
very slowly recovering damage after a proton fluence of up to
\expfor{5}{13}\,cm$^{-2}$.

\section{The crystals}\label{s-crystals}

For our studies we used PbWO\,$_4$ crystals produced by the
Bogoroditsk Techno-Chemical Plant (BTCP) in Russia for the
electromagnetic calorimeter (ECAL) of the CMS
experiment\,\cite{etdr}. The crystals have the shape of truncated
pyramids with nearly parallelepipedic dimensions of
$2.4\!\times\!2.4$\,cm$^2$ transversely and 23\,cm length.  All
crystals were of production quality, and only because of slightly
non-compliant mechanical dimensions or mild mechanical surface damage
they could not be used in the ECAL construction.  In particular, they
were all pre-characterised for their radiation hardness through
$\gamma$-irradiation at the Geneva Hospital $^{60}$Co facility
following a standard CMS qualification procedure. Throughout the paper
we will refer to the results of this pre-characterisation as
$\muindhosp$.

The studies presented here concentrated on an examination of the
change in longitudinal transmission (LT) at a wavelength of 420\,nm,
which corresponds to the peak emission of PbWO\,$_4$ scintillation
light. The CMS technical specifications\,\cite{note98:038} require the
radiation-induced absorption coefficient
\begin{equation}
\muindl = \frac{1}{\ell}\times \ln\frac{LT_0(\lambda)}{LT(\lambda)},
\label{muindeq}
\end{equation}
at $\lambda$=420\,nm to be below 1.5\,m$^{-1}$ for a saturated damage from 
$\gamma$-irradiation at $\approx$100\,Gy/h.
In Eq.\,\ref{muindeq} $LT_0$ ($LT$) is the longitudinal transmission
value measured before (after) irradiation and $\ell$ is the crystal
length (23\,cm) through which transmission is measured.

\begin{table}
\begin{center}
\begin{tabular}{c|c|c|l||c|c|r}
\multicolumn{4}{c||}{Crystals for proton irradiations}                             & \multicolumn{3}{c}{Crystals for $\gamma$-irradiations} \\ \hline
ID & $\muindhosp$              & $\phi_p$ (cm$^{-2}$h$^{-1}$) & $t_{\rm irr}$\,(h) & ID & $\muindhosp$  & $t_{\rm irr}$(h:min) \\ \hline
{\it a}  & 0.28                & $10^{12}$                    & 1 + 10 + 50        &{\it  t}  & 1.22            & 27:20            \\
{\it b}  & 0.79                & $10^{12}$                    & 1                  &{\it  u}  & 0.24            & 1:00             \\
{\it c}  & 0.90                & $10^{12}$                    & 1                  &{\it  v}  & 0.25            & 46:00            \\
{\it d}  & 0.39                & $10^{12}$                    & 10                 &{\it  w}  & 0.21            & 8:20             \\
{\it E}  & 0.70                & $10^{13}$                    & 1                  &{\it  x}  & 1.6             & 8:20             \\
{\it F}  & 0.55                & $10^{13}$                    & 0.1 + 1            &{\it  y}  & 1.14            & 46:00            \\
{\it G}  & 0.21                & $10^{13}$                    & 0.1                &{\it  z}  & 1.96            & 1:00             \\ \cline{5-7}
{\it h}  & 0.50                & \expfor{5}{11}               & 20                 & \multicolumn{3}{|c}{} \\ \cline{1-4}
\end{tabular}
\end{center}
\caption{List of all crystals, their initial radiation hardness
  pre-characterisation ($\muindhosp$ in m$^{-1}$) based on the
  standard CMS $\gamma$-irradiation
  procedure\,\protect\cite{etiennette}. $t_{\rm irr}$ is the exposure
  time in the proton beam with nominal flux $\phi_p$ or in the
  1\,kGy/h $\gamma$-facility, used in this work.}
\label{table1}
\end{table}

For the proton irradiations we had 12 crystals at our disposal. In
order to minimise the effect of crystal-to-crystal variations, we
selected for our irradiation 8 crystals which showed the most
consistent behavior in optical characteristics and in terms of
$\muindhosp$. The selected crystals are listed as {\it a--h} in
Table\,\ref{table1}.

For the $\gamma$-irradiations, performed at a dose rate corresponding
to $10^{12}$\,p/cm$^2$, we used further 6 crystals, from a later
production, which are listed as {\it u--z} in Table\,\ref{table1}.
The pre-characterisation of these 6 crystals was done for 1\,h at
350\,Gy/h which increases $\muindhosp$ by about
10\%\,\cite{etiennette} with respect to 2\,h at 250\,Gy/h, as used for
the other crystals.  Crystals {\it u, v} and {\it w} were of very good
quality while the three others showed after the pre-characterisation
quite high $\muindhosp$ values. Two of them ({\it x} and {\it z}) were
actually outside of the CMS specifications. In addition to these 6,
one of those remaining from the earlier 12, was considered good enough
to be used in the $\gamma$-irradiations and is labelled as {\it t} in
Table\,\ref{table1}.

The target dose rate in the $\gamma$-irradiations was 1--2\,kGy/h but
an exact analysis of the actual rate is presented later. Throughout
this paper we use the convention to assign small letters to crystals
irradiated at a dose rate of $\sim$1\,kGy/h and capital letters to
those irradiated at rates of $\sim$10\,kGy/h.  A prime or a
double-prime after the crystal identifier indicates a second or third
irradiation of the same crystal.

The $t_{\rm irr}$ times given in Table\,\ref{table1} for the proton
irradiations are the target values but the actual fluences were
determined by standard activation techniques. For the
$\gamma$-irradiations, performed in static and very stable conditions,
the $t_{\rm irr}$ values are the actual exposure times which are used
in the dose determination.

\section{Irradiations}

For the proton irradiations we used the IRRAD1 facility\,\cite{irrad1}
in the T7 beam line of the CERN PS accelerator, which at the time of
our first irradiations was delivering a proton beam of 20\,GeV/c and
during a later campaign 24\,GeV/c.  In order to obtain a uniform
irradiation over the whole crystal front face, we requested a special
beam, which had not been produced before in that facility. In
particular, we asked for the lowest beam intensity which the PS could
provide in a controllable way in the T7 beam line.

At 20\,GeV/c we obtained a flux of $\sim\!10^{12}$\,cm$^{-2}$h$^{-1}$
over an area of roughly $4.5\!\times\!4.5$\,cm$^2$. At 24\,GeV/c a
further intensity reduction by a factor of 2 was reached. The
uniformity was not perfect over the whole beam spot, but a fairly
uniform area covering the crystal could be found. In this
low-intensity mode the PS was delivering on average one proton spill
every 42\,seconds. The duration of the spill was 600\,ms at 20\,GeV/c
and 400\,ms at 24\,GeV/c.

In order to investigate a possible rate dependence, we performed some
irradiations at a higher intensity, of
$\sim\!10^{13}$\,cm$^{-2}$h$^{-1}$. To achieve this, the spill
frequency was increased to 2 spills every 14\,s and the beam spot was
reduced to a size of about $3\times 3$\,cm$^2$ which was still
sufficient to cover the crystal while maintaining an acceptable beam
uniformity.

The crystals were brought into the beam by a remotely controlled
shuttle. The beam was hitting the small front face of the crystal and
was parallel to the long crystal axis.

The $\gamma$-irradiations were performed at the Calliope plant of
ENEA-Casaccia, Italy. The facility provides a $^{60}$Co source which
at the time of our irradiations had a total activity of 750\,TBq. The
48 source rods were arranged in two concentric cylinders of about
20\,cm outer radius and 26\,cm height. We irradiated simultaneously up
to 5 crystals, standing side-by-side at a distance of 40\,cm from the
source centre, with their long side parallel to the source axis. Each
crystal was housed in its own styropor box. The distance between two
adjacent crystals was about 34\,mm. In order to obtain as uniform an
exposure as possible, each box was turned once by 180$^o$ around its
long axis at half-time through its exposure. Unfortunately the
crystals could not be precisely adjusted with respect to the source
and an offset of 52\,mm remained along the axis. Thus one extremity of
the crystals was slightly less exposed.

\section{Measurement methods}

\subsection{Light transmission}

The LT measurements were taken with a Perkin Elmer Lambda 900
spectrophotometer for wavelengths between 300 and 800\,nm in 1\,nm
steps. Since the grating which allows to select the desired
wavelengths polarises light, we used a wide-band depolariser to be
reasonably free of polarisation effects. The width of the beam is
defined by the slit opening which fixes the wavelength distribution:
this we set to $\Delta\lambda=1$\,nm which yields about 7\,mm beam
size. The height was adjusted with a ``common beam mask'' to
approximately 10 mm.

We will discuss later the implications of non-uniform proton beam
profiles, which required us to align the crystals in a reproducible
way. For each measurement we checked that the entry and exit points of
the light were centred on the crystal face.  The accuracy of this
alignment was roughly $\pm$1\,mm. Measurements with intentionally
larger displacements showed this to be sufficiently good not to
introduce significant variation for any of our crystals.

\begin{figure}[h]
\begin{center}
\includegraphics*[height=7cm]{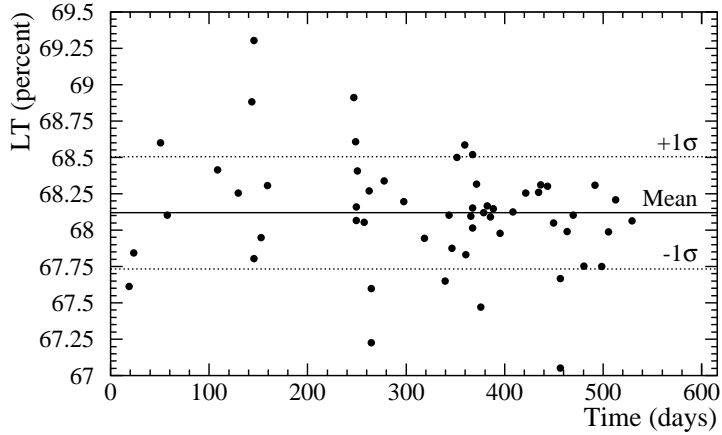}
\end{center}
\caption{LT data measured on the reference crystal at 420\,nm. The 0-point of 
the time axis is arbitrary, but the relative time intervals correspond to real
elapsed time.}
\label{fig2}
\end{figure}

For each proton-irradiated crystal, at least one set of LT
measurements was performed, where the light was passed not only
through the centre, but also as close to each of the four sides as
possible. The information gained from these ``scans'' served the
purpose of understanding effects of beam non-uniformities.

Before each session of measurements we experimentally verified the
absolute calibration of the device by measuring the LT of a
non-irradiated reference crystal. Fig.\,\ref{fig2} shows the results
of these measurements as a function of time at 420\,nm.  From the
scatter of the points we derive a 1$\sigma$ deviation of
0.58\%. However, close to the band-edge the transmission changes
rapidly and the accuracy of the wavelength dominates.  We were able to
reproduce the scatter of the reference crystal LT close to the band
edge, if we assumed that the wavelength is reproduced with 0.23\,nm
accuracy. Thus we estimate the total error of the LT measurement at
wavelength $\lambda$ as \be \Delta[LT(\lambda)] = 0.0058\times
LT(\lambda) \oplus 0.23\times \delta_{LT(\lambda)}, \ee where
$\delta_{LT(\lambda)}$ is the difference between two adjacent LT
points, 1\,nm apart.  The second term is significant only close to the
band edge, when $\delta_{LT(\lambda)}$ can be as big as three
percent-units per nm. At large wavelengths, when the transmission
changes slowly with $\lambda$, the error is given by the first term
alone.

\subsection{Induced radioactivity}

The induced radioactivity dose rate ($\iadr$) was measured with an
Automess 6150AD6 \cite{automess} at a distance of 4.5\,cm from the
long face of the crystal at its longitudinal centre. The device was
calibrated to yield the photon equivalent dose, which is the quantity
extracted from the simulations described in
appendix\,\ref{app-indact}.

The reference point of the sensitive element in the 6150AD6 is
reported to be 12\,mm behind the entrance window. Thus our actual
distance was 57\,mm from the crystal face. The $\iadr$ values of the
least active crystals were comparable to natural background, which we
determined by integrating over a long period with the 6150AD6.  In
order to verify the linearity of the device even at low counting
rates, we also measured the dose from a standard point-like $^{137}$Cs
source at various distances.  Both methods gave a background of
0.06\,$\mu$Sv/h and we observed no dependence on counting rate.

The nominal energy range of the 6150AD6 is 60\,keV--1.3\,MeV. Below
60\,keV the efficiency drops abruptly, but in our crystals only a
negligible fraction of dose comes from low energy photons.  The
efficiency curve\,\cite{automess}, which rises steeply above 1\,MeV,
is cut at 1.3\,MeV, which marks the end of the calibrated range. Our
assumption of constant efficiency above this energy most likely leads
to an underestimate of our simulations with respect to the measured
values.

Since PbWO$_4$ attenuates photons quite efficiently, a measurement of
$\iadr$ provides a means to approximately determine the proton beam
uniformity. For this purpose $\iadr$ was measured on all long sides
for each crystal.  Although the attenuation is exponential, a simple
average over the four sides gives the total activation with an
accuracy of a few percent even for our most non-uniform irradiations.

All $\iadr$ measurements were prolonged sufficiently to reduce the
statistical uncertainty to the level of 1--2\%.

\section{Simulation of the irradiation conditions}

\subsection{Proton beam}

\begin{figure}[h]
\begin{center}
\includegraphics*[height=13cm]{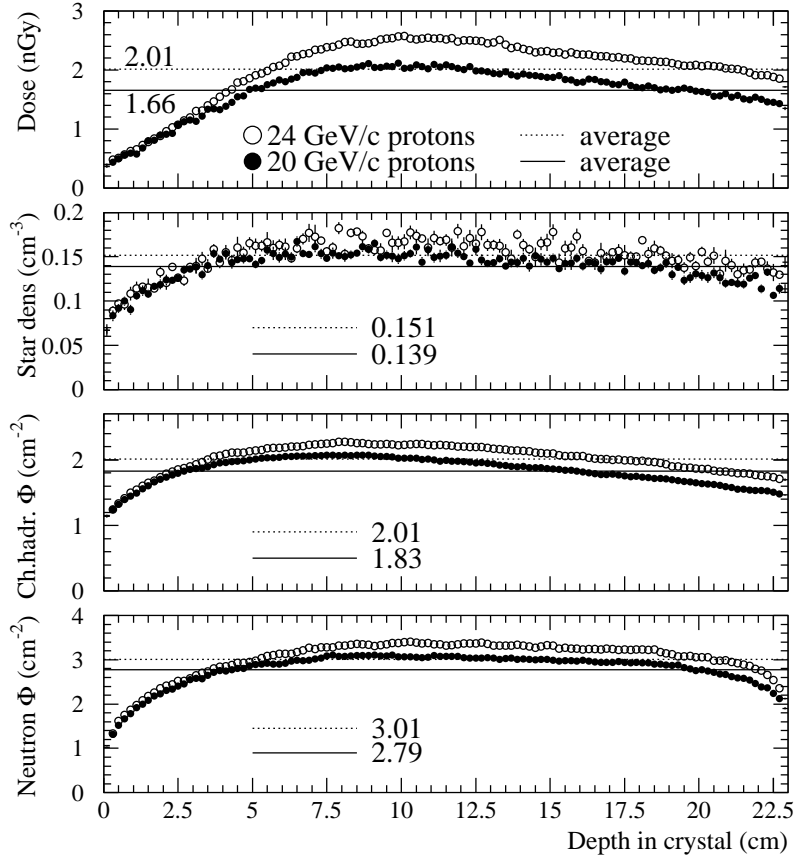}
\end{center}
\caption{Ionising dose, star density, charged hadron and neutron
  fluences per unit of incident proton fluence as obtained from the
  simulations. Values are for a 5$\times$5\,cm$^{2}$ proton beam of
  20\,GeV/c (solid dots and lines) or 24\,GeV/c (open dots, dashed
  lines). The lines and associated values give the average over the
  whole crystal length in the central area, covered by the light beam
  in the LT measurements.}
\label{fig3}
\end{figure}

Since the crystals are longer than one hadronic interaction length,
the protons initiate an intense hadronic cascade. In addition, the
irradiation zone is a fairly small volume with steel walls on 4 sides
at distance of few tens of centimetres. The back wall is in fact the
beam dump of the T7 beam line.  Thus the possible effects of
back-scattered radiation cannot be ignored.

In order to properly understand the radiation field within and around
the crystal -- and in particular the influence on the determination of
the incident proton fluence -- a full simulation of the irradiation
setup was performed with the FLUKA code\,\cite{fluka}. For the
simulations, the crystal shape was described as a 23\,cm long
truncated equal-sided pyramid with a minor face of
22$\times$\,22\,mm$^2$ and a major face of 26$\times$26\,mm$^2$.  The
main quantities characterising the radiation field, as obtained from
these simulations, are shown in Fig.\,\ref{fig3}.

\begin{figure}[h]
\begin{center}
\includegraphics*[height=7cm]{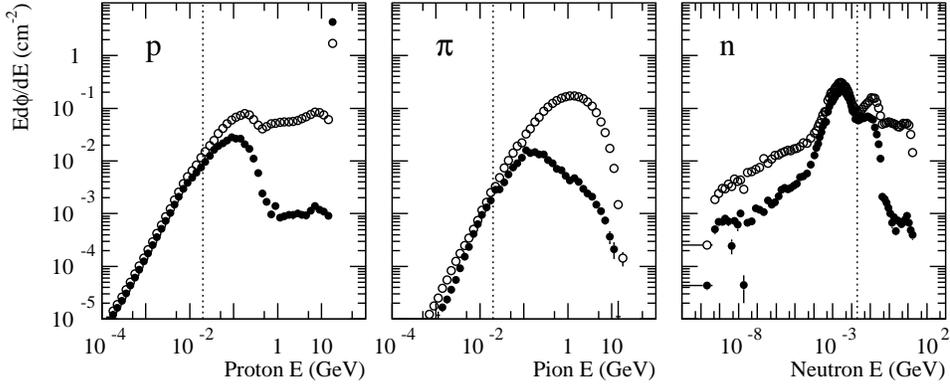}
\end{center}
\caption{Energy spectra of protons, charged pions and neutrons in the
  first (solid circles) and last (open circles) millimetre of a 23\,cm
  long crystal.  The dotted line shows the 20\,MeV threshold above
  which hadrons can produce stars.}
\label{fig4}
\end{figure}

\begin{table}
\begin{center}
\begin{tabular}{c|c|c|c|c}
       &Beam protons & $>$20\,MeV protons & $>$20\,MeV pions & $>$20\,MeV neutrons \\ \hline
Front  & 0.997       & 1.056              & 0.041            & 0.163               \\ 
Rear   & 0.390       & 0.764              & 0.497            & 0.536               \\ \hline
\end{tabular}
\end{center}
\caption{Simulated hadron fluxes (cm$^{-2}$ per unit of incident flux)
  in the first and last millimetre of the 23\,cm long crystal hit by a
  20\,GeV/c proton beam.}
\label{table2}
\end{table}

Damage caused by $\gamma$-irradiation is found to saturate at a level
corresponding to the dose rate used \cite{nim438:415}. However, since
we are looking for the possible existence of cumulative
damage\footnote{Earlier hadron irradiations reported in
  \protect\cite{note98:069,nim530:286} did not show any difference
  between $\gamma$ and pion irradiation performed at the same rate.
  But these tests cannot be considered sufficient to exclude possible
  cumulative effects because they were not extended to high enough
  fluences.} specific to hadrons, it seems more appropriate to use,
instead of dose rate, a purely hadronic quantity against which we
could plot the damage.

The incident proton beam is practically monoenergetic but, due to
cascade formation, the hadron spectrum within the crystal, shown in
Fig.\,\ref{fig4}, comprises all energies from sub-MeV neutrons to full
beam energy. Table\,\ref{table2} shows the integral fluences at beam
energy and above 20\,MeV -- the threshold indicated by the dashed
lines in Fig.\,\ref{fig4}.  With such a wide energy spectrum and since
a dependence between damage and hadron energy cannot be excluded {\it
  a priori}, it is not obvious how damage could be related to hadron
fluence.

It is plausible that any damage unique to hadrons should be caused by
the component of the hadron spectrum which is producing inelastic
interactions.  The most suitable quantity to describe the radiation
field for our purpose, is called {\em star density}, which was
originally introduced to parametrise radioactivity induced by hadron
irradiation\,\cite{stardens}.  A {\it star} is defined as an inelastic
hadronic interaction caused by a projectile above a given threshold
energy. This means that the star density is actually the integral of
the total hadron fluence above a given threshold, weighted by the
inelastic cross section. The star density is exactly zero for
irradiations with low energy photons or neutrons below the threshold
energy, which we take as 20\,MeV.  For constant beam conditions the
star density is of course proportional to the fluence. In our analysis
we will prefer the latter because it can be experimentally determined
with good accuracy. The simulated star densities, however, will be
used to scale between different beam energies and to compare different
radiation environments.

Since the radioactivity produced in the crystal is roughly
proportional to the rate at which stars are
produced\,\cite{note02:019}, a measurement of $\iadr$ in a crystal can
provide an estimate of the star density. This is discussed in more
detail in appendix\,\ref{app-indact}.

\subsection{$^{60}$Co source}
\label{cosim}

\begin{figure}[h]
\begin{center}
\includegraphics*[width=0.8\textwidth,bb = 0 240 680 680, clip]{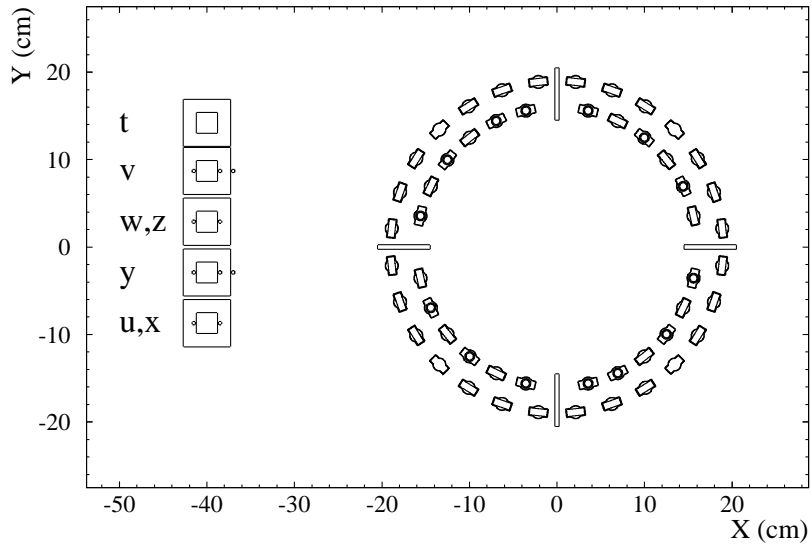}
\end{center}
\caption{Top view of the simulation geometry representing the Calliope
  facility. The two concentric assemblies of source rod rings are seen
  on the right and the 5 crystals inside their boxes on the left. The
  positions of each crystal are indicated by their identifiers next to
  the boxes. The small dots on and within the crystal boxes represent
  alanine dosimeter positions.}
\label{fig5}
\end{figure}
\begin{figure}[b]
\begin{center}
\includegraphics*[height=10cm]{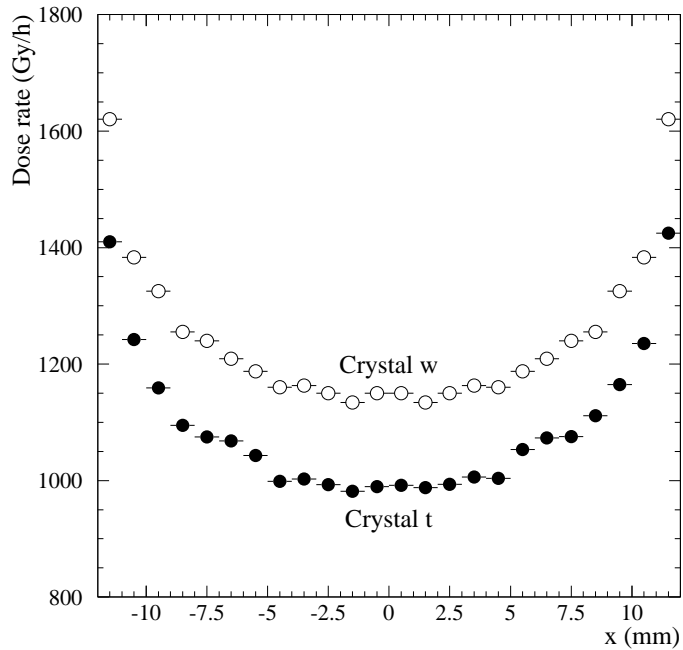}
\end{center}
\caption{Simulated transverse dose rate profiles in the
  crystals. Shown are the minimum (crystal {\it t}) and maximum
  (crystal {\it w}) dose rates in the 7 crystals.  The profiles
  represent an average over the whole length of the crystal.}
\label{fig6}
\end{figure}
\begin{figure}[t]
\begin{center}
\includegraphics*[height=10cm]{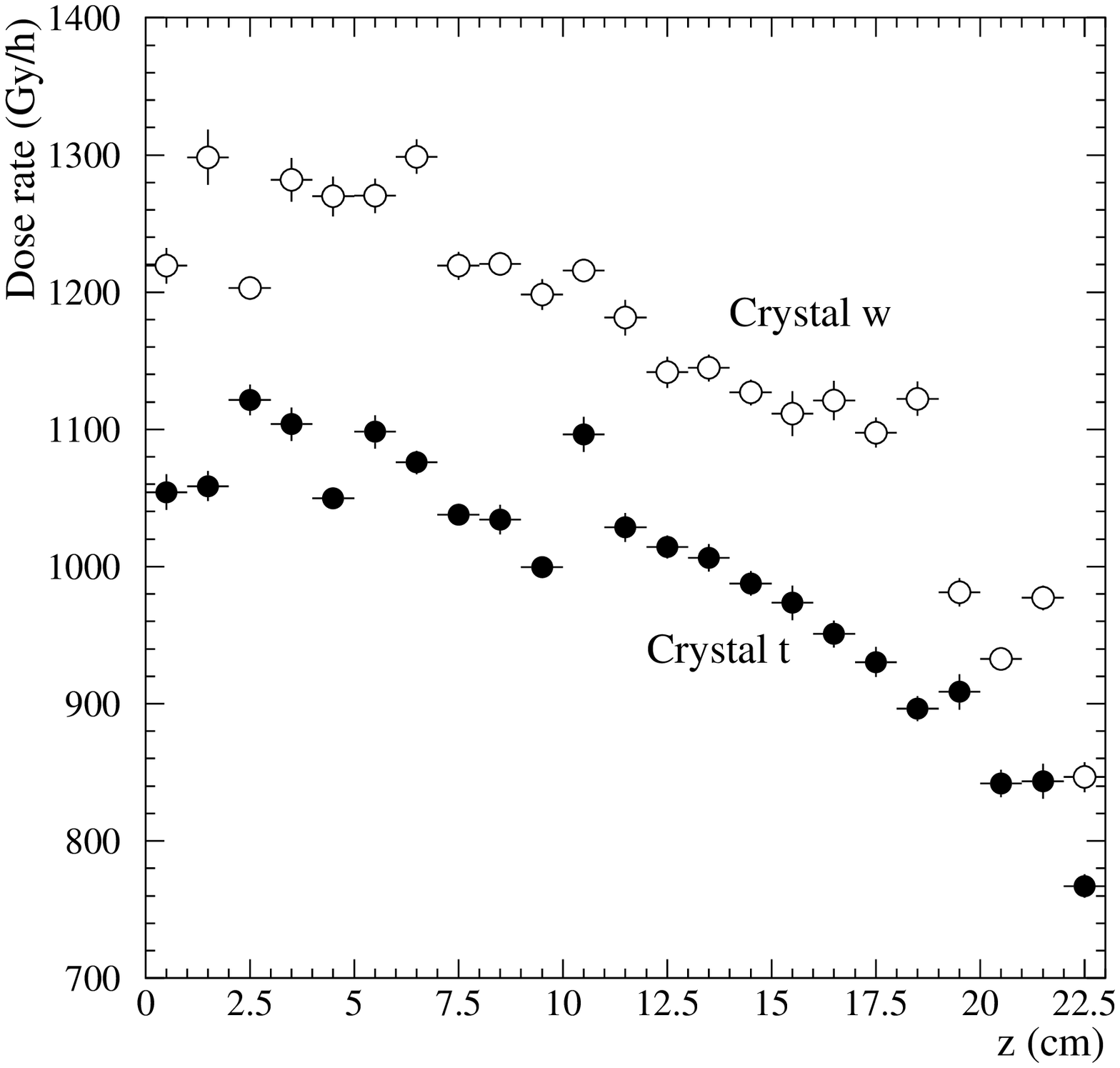}
\end{center}
\caption{Simulated longitudinal dose rate profiles along the centre of the crystals ($\pm$5\,mm around central axis). 
Shown are the minimum (crystal {\it t}) and maximum (crystal {\it w}) dose rates in the 7 crystals.}
\label{fig7}
\end{figure}
The dose rates at the Calliope plant\,\cite{tn95:192} are well known from measurements, performed
regularly by the operators. However, these values refer to dose in air or water in the absence 
of scattering material. In order to determine the dose profile within the crystals, detailed 
simulations of the irradiation setup were mandatory. Fig.\,\ref{fig5} shows the FLUKA simulation 
model of the source and 5 crystals in their boxes. Since durations of irradiations
were different, not all positions were occupied at all times. Due to these changes in
configuration and the slightly different position with respect to the source, the dose 
in individual crystals is slightly different.

As in the case of proton irradiated crystals, we again determine the transmission along the long axis of
the crystal, and therefore we are interested in the dose along the centre. Fig.\,\ref{fig6} shows
the simulated transverse dose rate profiles, which exhibit a pronounced U-shape, caused by turning the crystal
half-way through the irradiation. The fairly flat central part, of about $\pm$5\,mm width,
fits very well the width of the light beam in the LT-measurements. In the other dimension the 
profile is flat. 

Figure\,\ref{fig7} shows the simulated longitudinal dose rate profile along the long axis of the 
crystal. The dose drops quite significantly towards the end of the crystal because 
the height of the source could not be exactly aligned with the crystals. The discontinuities, e.g. the
step at 18\,cm for crystal {\it w}, are caused by the support structures of the source assembly.

\section{Fluence and dose determination}

\subsection{Proton fluence determination in IRRAD1}

The fluence, $\Phi_p$, of the incident proton beam was determined with
aluminium foils, cut to a size of $24\times 24$\,mm$^2$ and placed in
front of the crystals at a distance of about 10\,mm. Secondaries
produced by the crystals, neutrons in particular, influence this
measurement. Therefore we preferred to use the isotope $^{22}$Na
instead of the more standard $^{24}$Na. The determined fluence, in
particular its relation to the star density within the crystal, is
sensitive to the geometry and alignment. A further factor to be taken
into account came from a change of beam momentum from the initial
20\,GeV/c to 24\,GeV/c.  Therefore we paid great attention to a
careful fluence analysis and cross checks with the induced
radioactivity in the crystal itself. The technical aspects of this and
all effects considered, together with the correction factors derived
by us, are detailed in appendix\,\ref{app-pflux}.

\begin{table}
\begin{center}
\begin{tabular}{c||c|c||c||c}
ID & $\Phi_p$($^{22}$Na) (cm$^{-2}$) &  $\sum\Phi_p$ (cm$^{-2}$)& Dose rate (kGy/h) & Dose (kGy) \\  \hline
b  & \expfor{(0.46\pm 0.03)}{12} &  \expfor{(0.46\pm 0.03)}{12} & 0.99              & 0.77       \\ 
c  & \expfor{(0.74\pm 0.05)}{12} &  \expfor{(0.74\pm 0.05)}{12} & 1.40              & 1.23       \\ 
a  & \expfor{(1.01\pm 0.11)}{12} &  \expfor{(1.01\pm 0.11)}{12} & 1.06              & 1.67       \\ 
F  & \expfor{(1.37\pm 0.16)}{12} &  \expfor{(1.37\pm 0.16)}{12} & 22.8              & 2.28       \\ 
G  & \expfor{(1.72\pm 0.18)}{12} &  \expfor{(1.72\pm 0.18)}{12} & 28.5              & 2.85       \\ 
E  & \expfor{(7.66\pm 0.55)}{12} &  \expfor{(7.66\pm 0.55)}{12} & 15.9              & 12.7       \\ 
h  & \expfor{(8.70\pm 1.00)}{12} &  \expfor{(9.46\pm 1.09)}{12} & 0.66              & 17.4       \\ 
F' & \expfor{(8.46\pm 0.85)}{12} &  \expfor{(9.83\pm 0.86)}{12} & 18.0              & 16.3       \\ 
d  & \expfor{(12.0\pm  1.2)}{12} &  \expfor{(12.0\pm  1.2)}{12} & 1.74              & 20.0       \\ 
a' & \expfor{(12.2\pm  1.0)}{12} &  \expfor{(13.2\pm  1.1)}{12} & 1.66              & 21.9       \\ 
a" & \expfor{(37.6\pm  2.3)}{12} &  \expfor{(54.1\pm  2.7)}{12} & 1.25              & 97.5       \\ \hline
\end{tabular}
\end{center}
\caption{Proton fluences $\Phi_p$($^{22}$Na) for each individual
  irradiation as obtained from the Al-activation and cumulative
  fluences ($\sum \Phi_p$) for a given crystal. The latter also
  includes the correction for beam energy (see text).  The last two
  columns give the average ionising dose rate and integral dose along
  the axis of the crystal. The dose values are based on simulations,
  renormalised to the measured fluence at proper beam energy.}
\label{table3}
\end{table}

Table\,\ref{table3} summarises the proton fluences and gives the dose
rates and total doses averaged along the axis of the crystal. The
first column, $\Phi_p$($^{22}$Na), gives the fluence value obtained
from the aluminium foils but corrected by the factor of 0.89 in order
to account for activation by secondaries. The $\sum\Phi_p$-column
gives the integral of all subsequent irradiations for a given crystal
and also includes the correction by a factor of 1.086 to account for
the higher beam energy in the case of crystals {\it a"} and {\it
  h}. Any quantity that is cumulative and stable, should appear
proportional to $\sum\Phi_p$.

The dose values given in Table\,\ref{table3} correspond to an average
along the axis of the crystal, i.e.  along the path of light in the LT
measurements. Averaging was done over the dimensions of the light
beam. It can be seen from Fig.\,\ref{fig3} that in the proton
irradiations the maximum dose rate, at a depth of about 10\,cm within
the crystal, is $\sim$30\% above this average.

One issue which will be discussed in detail in
appendix\,\ref{app-nonunif} is the uniformity of the beam spot across
the crystal face. These effects, however, remained small enough to
require no corrections.  Actually, as shown in
appendix\,\ref{app-nonunif}, the simulated effects of non-uniform
irradiation agree nicely with measured variations of $\iadr$ and LT
close to the edges of the crystal.

\subsection{Dose determination at the $\gamma$-facility}

\begin{figure}[b]
\begin{center}
\includegraphics*[height=8cm]{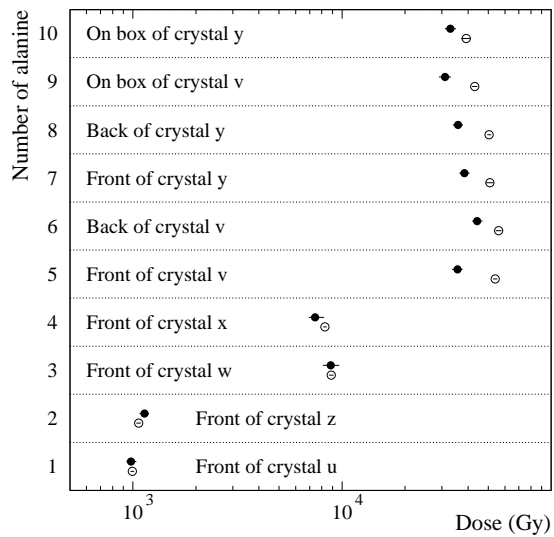}
\end{center}
\caption{Simulated (open circles) and measured (solid circles)
  cumulative doses in the 10 alanine dosimeters exposed during our
  $\gamma$-irradiations.}
\label{fig8}
\end{figure}

Alanine dosimeters provide a well-established method to determine
cumulative ionising doses.  The absolute calibration is usually done
with a standard $^{60}$Co source and converted to give the dose in
air. Ten alanine dosimeters which accompanied the irradiations
provided a crucial cross-check of our dose simulations.
\begin{table}[t]
\begin{center}
\begin{tabular}{c||c|c}
ID & Dose rate (kGy/h) & Dose (kGy)     \\ \hline
u  & 1.02        &  1.02  \\
z  & 1.15        &  1.15  \\
x  & 1.02        &  8.49  \\
w  & 1.15        &  9.59  \\
t  & 1.00        &  27.3 \\
y  & 1.06        &  48.9 \\
v  & 1.09        &  50.3 \\ \hline
\end{tabular}
\end{center}
\caption{Dose rates and cumulative doses reached in the
  $\gamma$-irradiations. Values are averages over the crystal length
  along the centre axis. The statistical errors are smaller than 1\%.}
\label{table4}
\end{table}
Six of these dosimeters were fixed on the ``front'' of each of the
crystals {\it u--z}, i.e. the side facing the source, while two were
on the ``back'' of crystals {\it v} and {\it y}. When turning the
crystals, ``front'' and ``back'' got inverted so that each of these
eight alanines was averaging equally over the ``front'' and ``back''
of a crystal.  A further two dosimeters were fixed on the source side
of the styropor boxes housing crystals {\it v} and {\it y}. These
dosimeters were removed when the boxes were turned after the first 23
hours of exposure.  All 10 alanines were included in the simulations
and the dose in them was recorded. Figure\,\ref{fig8} shows a
comparison between the doses determined from the alanines and the
simulated values.  While the low doses of the 1\,h and 8.3\,h
irradiations are in prefect agreement, alanines exposed for the
longest times indicate a lower value than the simulations. Due to the
symmetry of the simulated setup -- essentially the same result was
just rescaled with time -- it is inconceivable that the simulation
would fail for higher doses. The response of alanine is usually quoted
to be linear up to about 100\,kGy. However, when approaching the upper
limit, saturation sets in gradually. Although saturation is taken into
account in the calibration\,\cite{helmut}, an explanation for the
difference observed could still be if saturation depends also on the
photon energy spectrum, which is softer in the presence of the crystal
than for a plain $^{60}$Co calibration source.

Table\,\ref{table4} summarises the doses and dose rates reached in the
$\gamma$-irradiations.  These values have been extracted directly from
the simulation described in section\,\ref{cosim} but their accuracy
has been experimentally verified by the comparison with the alanine
data, shown in Fig.\,\ref{fig8}. The dose rate at one end was about
10\% higher than the average and dropped to 20\% below average at the
other end of the crystals, as can be seen in Fig.\,\ref{fig7}.
			      
\section{Results}
\begin{figure}[b]
\begin{center}
\includegraphics*[width=0.9\textwidth]{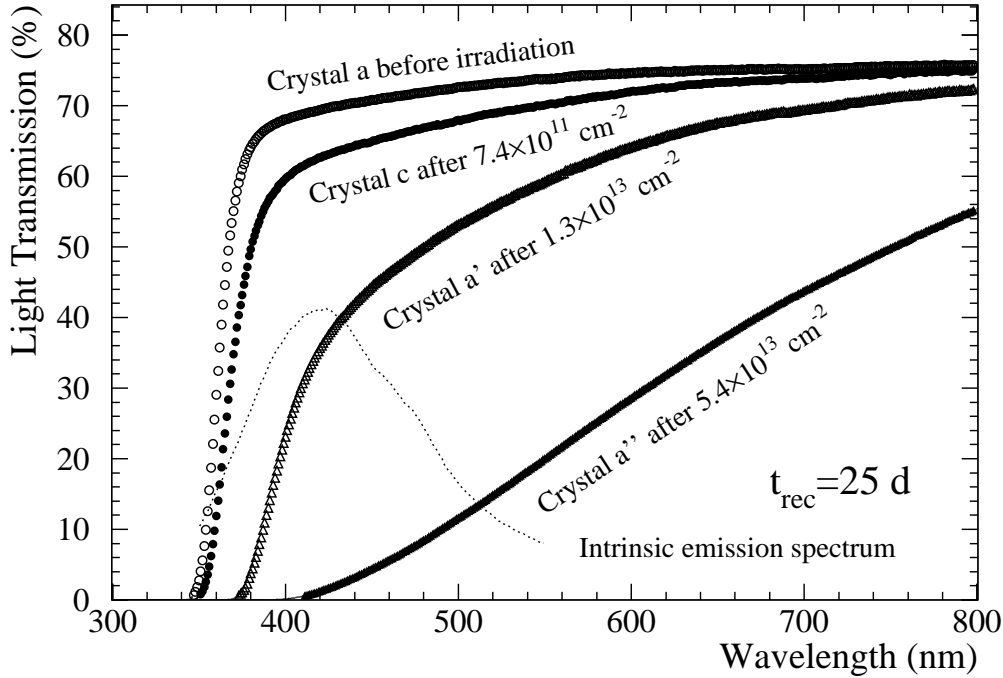}
\end{center}
\caption{LT curves for crystals with various degrees of proton induced
  radiation damage.  The emission spectrum (dotted line) is taken from
  \protect\cite{renyuan} and has arbitrary normalisation.}
\label{fig9}
\end{figure}
Fig.\,\ref{fig9} shows typical LT curves of a non-irradiated crystal
and of some crystals which have suffered radiation damage in the
proton beam.  With increasing $\sum\Phi_p$, not only the transmission
worsens, especially in the blue region of the spectrum, but in
addition the LT band edge shifts towards higher wavelengths. At
$\sum\Phi_p\!\approx\!10^{13}$\,cm$^{-2}$ this shift starts to
increasingly cut into the intrinsic emission spectrum of the crystal.
\begin{figure}[thb]
\begin{center}
\includegraphics*[height=9.5cm,bb=0 50 567 567]{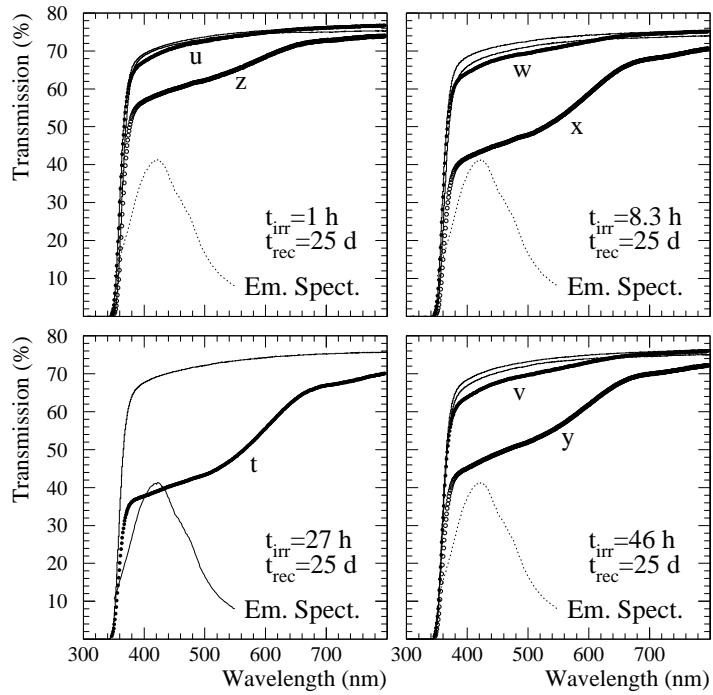}
\end{center} 
\caption{LT curves for crystals with various degrees of
  $\gamma$-induced radiation damage. The lines show the light
  transmission prior to irradiation. The order of these curves is the
  same as of the post-irradiation ones. The emission spectrum (dotted
  line) is taken from \protect\cite{renyuan} and has arbitrary
  normalisation.}
\label{fig10}
\end{figure}
\begin{figure}[bht]
\begin{center}
\includegraphics*[height=9.5cm,bb=0 50 567 567]{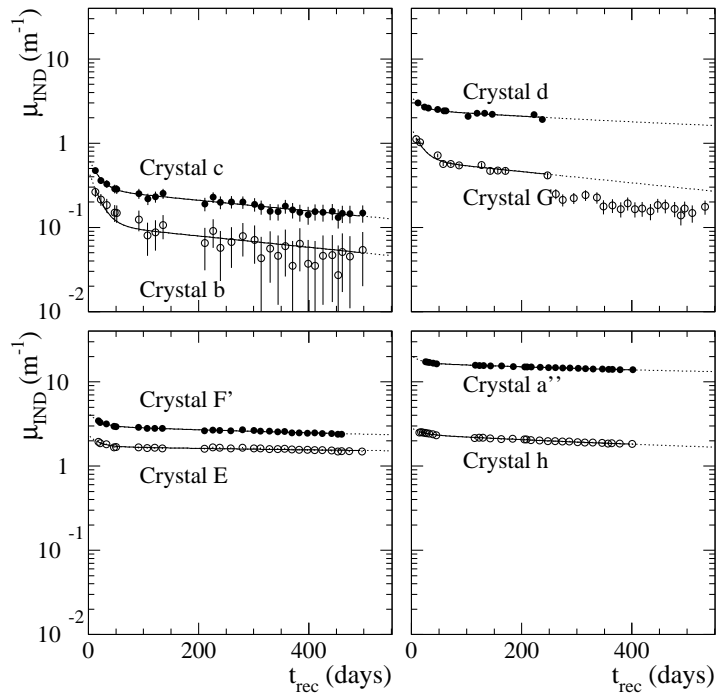}
\end{center}
\caption{The fits of Eq.\,\protect\ref{timefit} to recovery data of
  our proton irradiated crystals.  The last points of crystal {\it G}
  are excluded from the fit -- see text.}
\label{fig11}
\end{figure}
\begin{figure}[thb]
\begin{center}
\includegraphics*[height=9.5cm,bb=0 50 567 567]{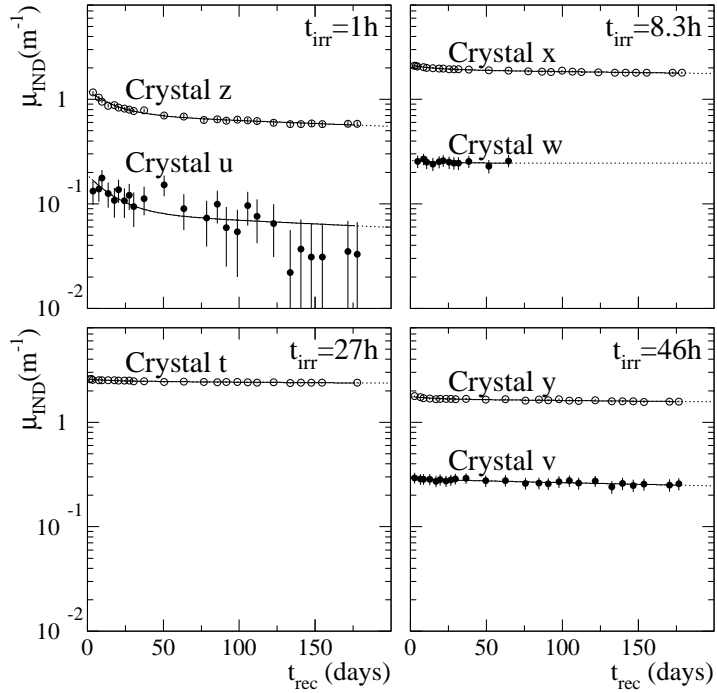}
\end{center}
\caption{Recovery data of crystals irradiated with $^{60}$Co photons
  up to various cumulative doses at a dose rate of $\sim$1.1\,kGy/h.}
\label{fig12}
\end{figure}
\begin{figure}[thb]
\begin{center}
\includegraphics*[width=0.9\textwidth,bb= 0 170 567 567]{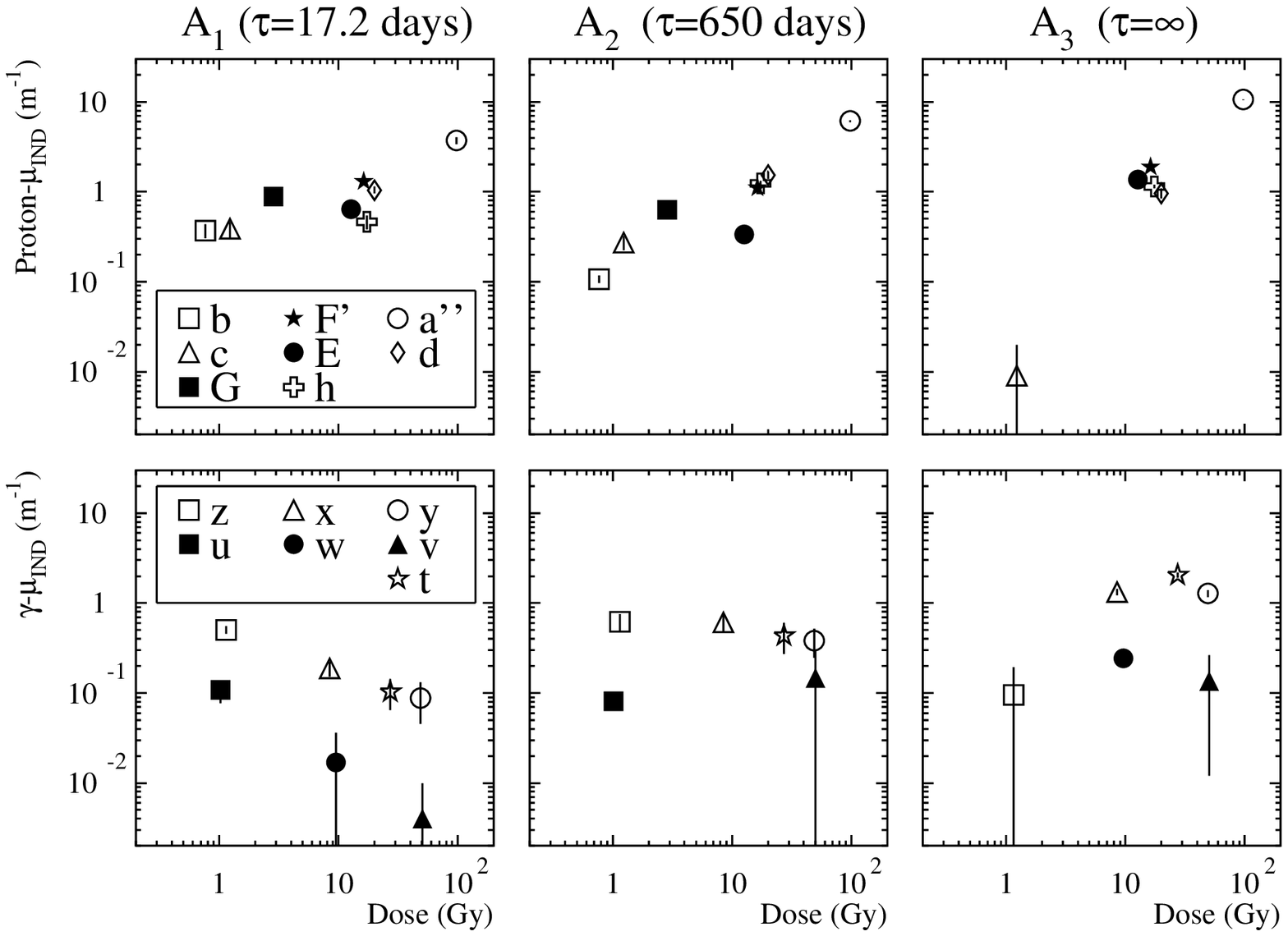}
\end{center}
\caption{The amplitudes $A_i$ of Eq.\,\protect\ref{timefit} as a
  function of dose. The dose to fluence correspondence for the
  proton-irradiated crystals can be found in
  Table\,\protect\ref{table3}. When symbols are not present in the
  plot, the fitted amplitude for that crystal is consistent with
  zero.}
\label{fig13}
\end{figure}

Fig.\,\ref{fig10} shows the transmission curves as a function of
wavelength for all 7 crystals irradiated with $^{60}$Co photons.  The
shapes of the LT curves of $\gamma$-irradiated crystals are completely
different from those of proton-irradiated ones. Even after the highest
cumulative dose reached, the band-edge in $\gamma$-irradiated crystals
does not shift at all, thus establishing a fundamental difference
between the damage caused by protons and the one caused by purely
ionising radiation. Instead, one recognizes the absorption band around
420\,nm, which is typical for $\gamma$-irradiated BTCP
crystals\,\cite{ieee47:1741}.  Although this band is probably also
present in proton-irradiated crystals, it does not appear in the LT
curves as a main feature of the damage.  Furthermore, in
$\gamma$-induced damage we see a significant crystal-to-crystal
variation.  This agrees well with the variation of the $\muindhosp$
values in Table\,\ref{table1}.  No significant correlation of this
kind is seen in the proton-irradiated crystals.

\subsection{Recovery time constants}
\begin{figure}
\begin{center}
\begin{tabular}{cc}
\includegraphics*[height=8cm]{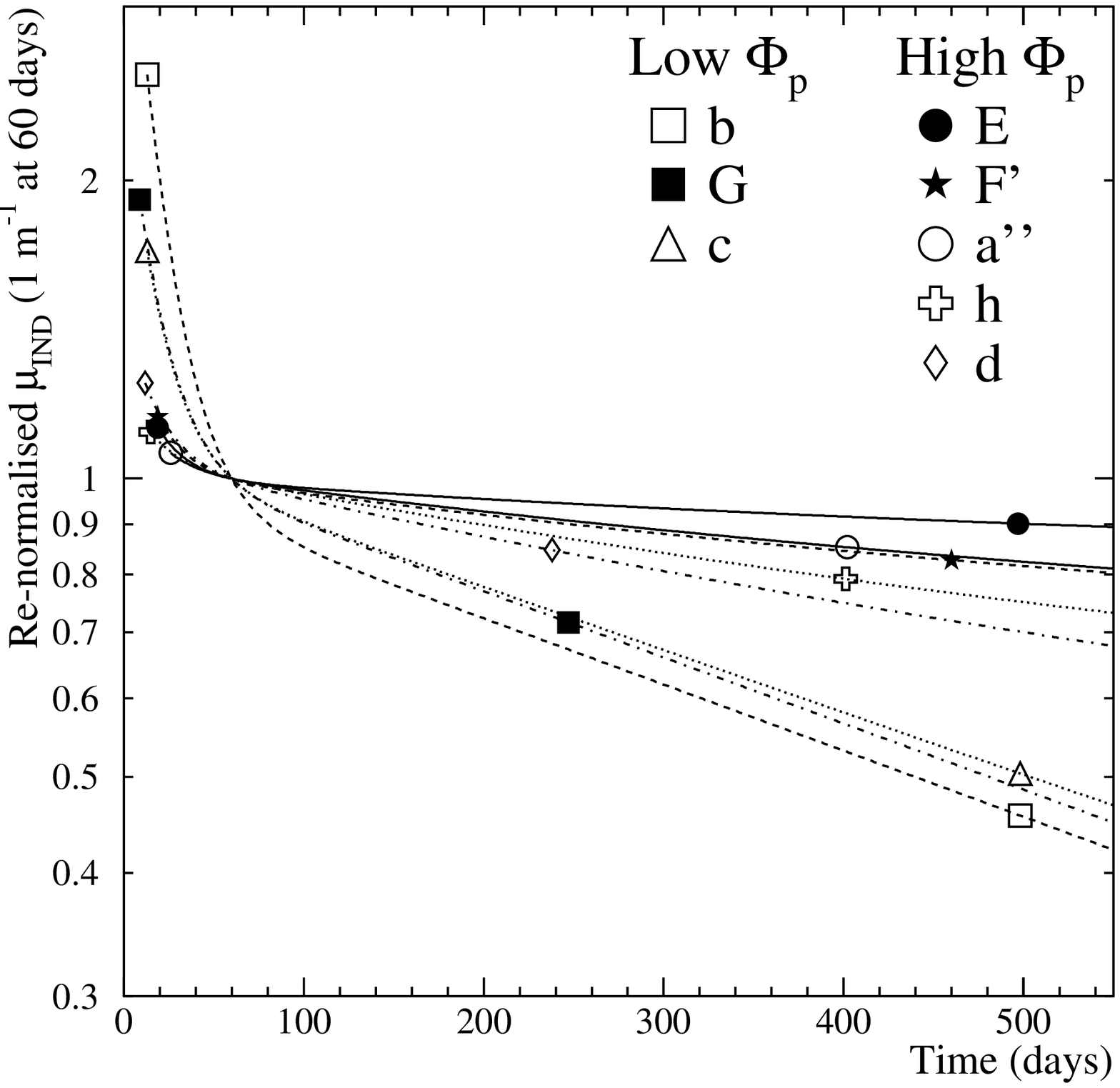}&
\includegraphics*[height=8cm]{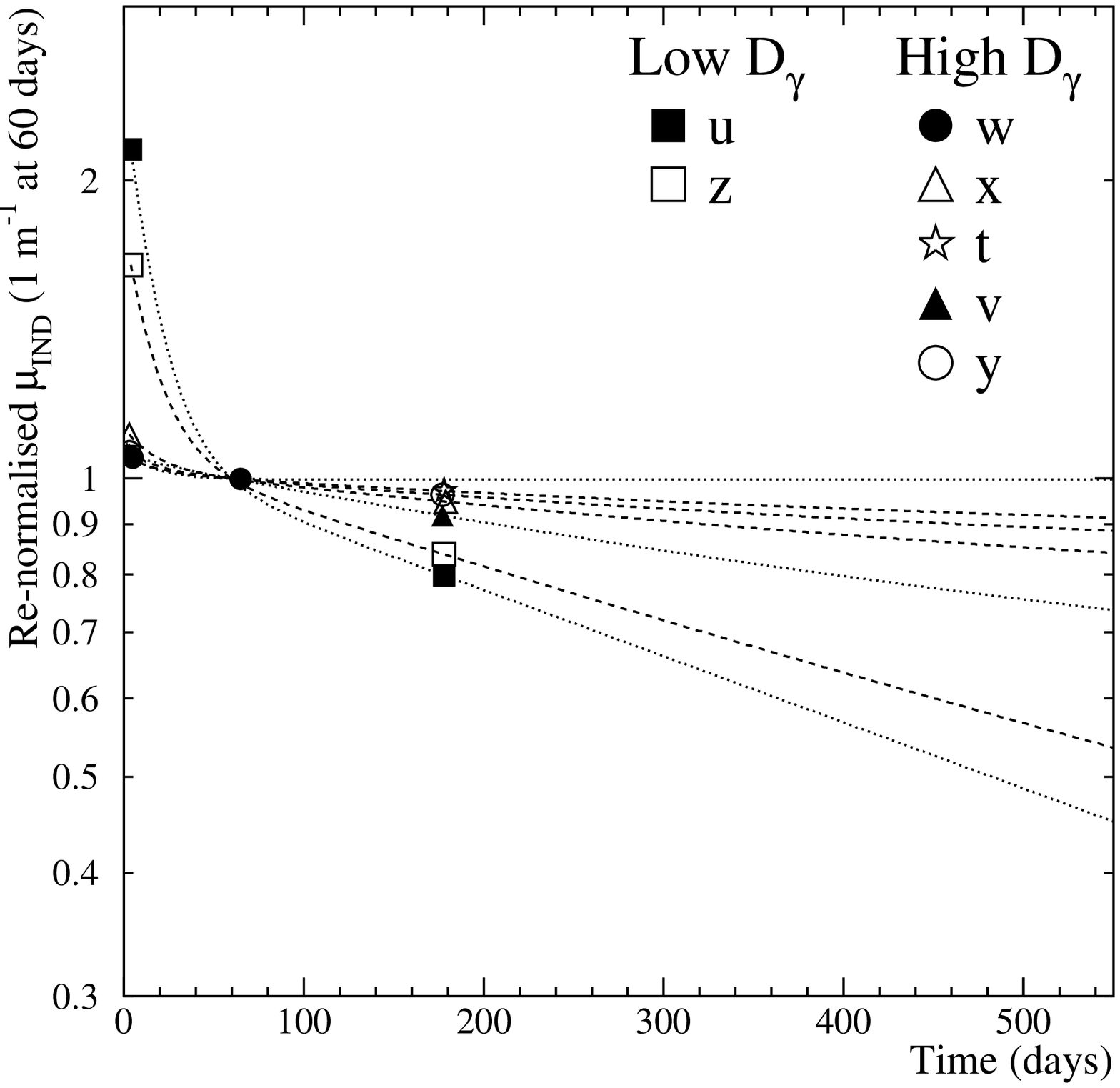}
\end{tabular}
\end{center}
\caption{Plots of Eq.\,\protect\ref{timefit} with the parameter values
  shown in Fig.\,\protect\ref{fig13} after renormalisation to unity at
  t=60\,days. The symbols indicate the range over which data were
  available for a given crystal.}
\label{fig14}
\end{figure}
We fit the time dependence of $\mu_{\rm IND}$ with a sum of two
exponentials and a constant
\be
\muindt=\sum_{i=1}^{2} A_i\exp\left(-\frac{t_{\rm rec}}{\tau_i}\right) + A_3,
\label{timefit}
\ee
where $t_{\rm rec}$ is the time elapsed since the
irradiation. Attempts to fit Eq.\,\ref{timefit} for each crystal, with
all 5 parameters free, revealed that the longer time constant
($\tau_2$) was comparable to, or larger than, $t_{\rm rec}$ and thus
the correlation between the second term and the constant was too
strong to allow for a reliable fit. It is, however, reasonable to
assume that the damage corresponds to the same color centres, each of
which corresponds to a unique time constant. Under this assumption we
fitted the $\tau_i$ for all crystals simultaneously and, with these
fixed, the amplitudes $A_i$ for each crystal separately. This process
was iterative, i.e. $\tau_i$ and $A_i$ were fitted repeatedly with
increasingly better initial guesses until convergence was reached. The
quality of the fits, in terms of $\chi^2$, was comparable to that
obtained by the 5-parameter fit of Eq.\,\ref{timefit} for each crystal
separately.

Figs.\,\ref{fig11} and \ref{fig12} show the fits for proton and
$\gamma$-irradiated crystals, respectively. In comparison to the
observed recovery time constants, crystals {\it a}, {\it a'} and {\it
  F} were re-irradiated too soon and therefore do not allow to obtain
reasonable fits.

Of the proton irradiated crystals, {\it d} was annealed 240 days after
irradiation, thus we have no longer-term room temperature recovery
data for it. While essentially no recovery of crystal {\it d} was
observed at 160\,$^o$C, partial recovery took place at
250\,$^o$C. This temperature was sufficient to restore the
pre-irradiation transmission at $\lambda\!>$400\,nm. At a temperature
of 350\,$^o$C almost complete recovery was reached, except for a small
residual shift of the band-edge.

In the case of crystal {\it G} a peculiar step appeared after 250
days, which we traced back to an unintentional exposure to normal room
light from fluorescent tubes for about 60\,hours during a long
radioactivity measurement. The existence of such light-induced
annealing was qualitatively verified by exposing to the same room
lights a crystal that had been proton irradiated while exercising the
procedures for this work. The very severe damage in that crystal had
remained stable for almost 2\,years but about 15\% of it annealed
fairly rapidly under light exposure.

The $\gamma$-irradiated crystal {\it w} was annealed 65 days after
irradiation, in preparation for a later proton irradiation.

From the global fit we obtained $\tau_1$=\tauyks\,days and
$\tau_2$=\taukaks\,days. The amplitudes corresponding to these
recovery time constants are shown in Fig.\,\ref{fig13}. The amplitude
corresponding to $\tau_1$ is slightly increasing with dose in the case
of proton irradiation while it decreases as a function of
$\gamma$-dose. This decrease could be due to the longer-lived defects,
which build up and contribute to light attenuation, making $A_1$
relatively less effective. Since $\tau_2$ was fitted to a value of
\taukaks\,days, which is longer than the period of the recovery
follow-up, the amplitudes $A_2$ and $A_3$ could not be reliably
resolved from each other. This was particularly true for all
$\gamma$-irradiated crystals and proton irradiated crystals $d$ and
$G$, where recovery data extends only to $t_{\rm
  rec}\sim$200\,days. However, the sum of $A_2+A_3$ for the proton
irradiated crystals increases linearly with fluence, while it is
consistent with a constant for $\gamma$-irradiated crystals.

All fits -- re-normalised to the same values at $t_{\rm rec}$=60\,days
-- are shown in Fig.\,\ref{fig14}. Re-normalisation at 60\,days was
chosen such that the $\tau_1$-component of damage has practically
disappeared. Thus the left side of the re-normalisation point shows
the relative amplitude of the damage corresponding to $\tau_1$, while
the right side provides a visual comparison of the magnitude of the
long-term damage.  Especially in terms of the short component the
crystals group according to the integral fluence or dose they
received. Those exposed to $\sum\Phi_p\!\sim\!10^{12}$\,cm$^{-2}$ (or
$\sim\!$1\,kGy) recover, in relative terms, more than those irradiated
to higher fluence.  While a slight difference between proton- and
$\gamma$-irradiated crystals can be seen in the magnitude of the fast
component, the long term recovery appears to be independent of the
radiation type.

\subsection{Correlation between light transmission and $\sum\Phi_p$ or $\gamma$-dose}
\begin{figure}[bht]
\begin{center}
\includegraphics*[height=10cm]{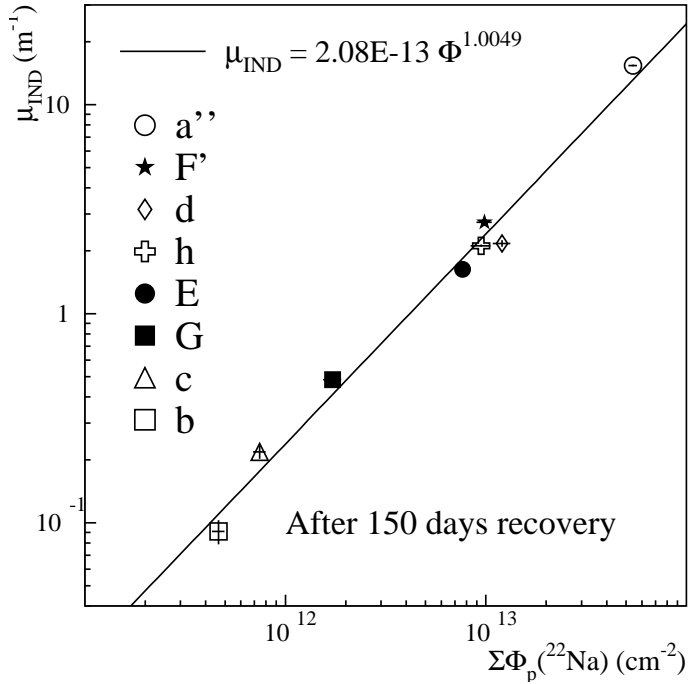}
\end{center}
\caption{Induced absorption $\muind$ as a function of cumulative proton fluence.}
\label{fig15}
\end{figure}
Figure\,\ref{fig15} shows $\muind$ as a function of fluence 150\,days
after irradiation as calculated from Eq.\,\ref{timefit} using the
parameter values shown in Fig\,\ref{fig13}.

For crystals irradiated several times, the cumulative fluence
$\sum\Phi_p$ has been used and $\muind$ has been normalised to the
transmission before the first irradiation.  We have fitted the data,
which cover two orders of magnitude, on a log-log scale and obtain an
exponent that corresponds to only 1\% deviation from linearity per
order of magnitude in fluence. In particular, if the data in
Fig.\,\ref{fig15} are compared with Table\,\ref{table1}, no
correlation between proton-induced damage and $\muindhosp$ can be
observed.
\begin{figure}
\begin{center}
\includegraphics*[height=13cm]{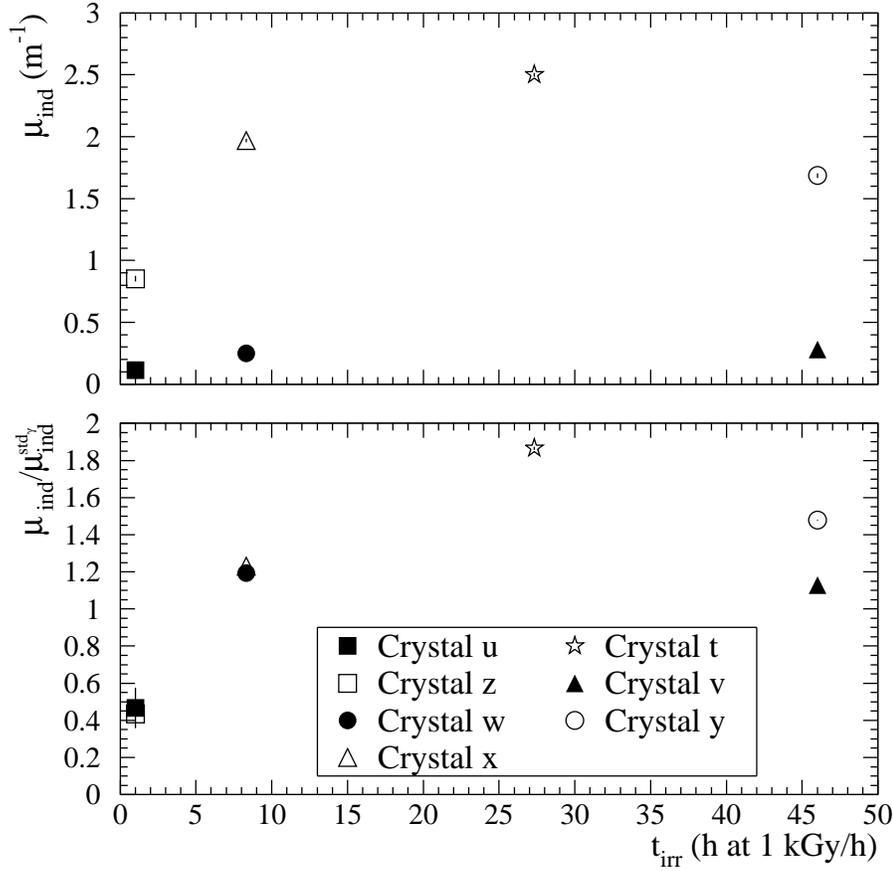}
\end{center}
\caption{Values of $\muind$ measured on the 7 $\gamma$-irradiated
  crystals after 20\,days recovery at room temperature. The upper plot
  shows the measured data and the lower one the ratio to the
  $\muindhosp$-values from Table\,\protect\ref{table1}.}
\label{fig16}
\end{figure}

Fig.\,\ref{fig16} confirms that even at a dose rate as high as
1\,kGy/h the $\gamma$-damage saturates. After 1 hour at 1\,kGy/h,
$\muind$ has not yet reached its plateau value, but between 8.3 and 46
hours the differences observed can be entirely attributed to
crystal-to-crystal variation. This becomes particularly clear when we
normalise $\muind$, as measured by us, with the $\muindhosp$-values
from Table\,\ref{table1}, which is supposed to cancel any
crystal-to-crystal variation. It should be noted, however, that the
$\muindhosp$ values are obtained only about 1\,hour after irradiation,
i.e.  before short-term recovery is completed.  In fact, it has been
observed that about 60\% of the damage contributing to $\muindhosp$
recovers during the first 20\,days while the rest is much more
stable\,\cite{etiennette}. If this is taken into account in the
$\muind/\muindhosp$ ratio of Fig.\,\ref{fig16}, the 1\,h points are
nearly consistent with unity while the ratios for longer irradiations
get pushed to a value of about 3. This large ratio is at least partly
-- maybe even mostly -- due to the fact that the pre-characterisation
irradiations did not last long enough to saturate all color centres.

Crystal {\it t} shows slightly higher $\muind$ values than the other 6
crystals in Fig.\,\ref{fig16}.  The deviation is even more pronounced
in the $\muind/\muindhosp$ ratio. These differences might well be,
because {\it t} is from the earlier set of 12 crystals received for
the proton tests.  These had undergone a different
pre-characterisation procedure almost a year earlier. While the 10\%
difference in $\muindhosp$, due to the different dose rate, has been
taken into account, there could still be other systematic or
procedural effects that might have led to a slightly lower
$\muindhosp$ of crystal {\it t}.

When comparing the data of Figs.\,\ref{fig15} and \ref{fig16}, it
should be taken into account that the proton irradiated crystals all
had $\muindhosp<1$\,m$^{-1}$ and most of them should be compared with
the solid symbols in the upper plot of Fig.\,\ref{fig16}. In
particular, as can be seen from Table\,\ref{table1}, crystal {\it a"}
was characterised as having the same, excellent, radiation hardness as
crystal {\it v}. After exposure to comparable integral doses, the
proton irradiated crystal {\it a"} shows an induced absorption of
15\,m$^{-1}$ while the $\gamma$-irradiated crystal {\it v} has
saturated at a value of only 0.3\,m$^{-1}$.

The shape of the LT curves, in particular the shift of the band-edge
after proton exposure and the fact that two universal time constants
fit the data well but give very different dose-dependence for the
amplitudes in Eq.\,\ref{timefit}, leads us to the following
interpretation:
\begin{itemize}  
\item The damage corresponding to the two $\tau_i$ values is produced
  by proton and $\gamma$-irradiation in the same way and results in
  the same recovery-characteristics.  Most likely this damage
  corresponds to the same color centres.
\item The band-edge shift is unique to proton irradiation and
  corresponds to a different damage mechanism. This damage appears to
  be stable and cumulative, at least over the $\sum\Phi_p$ range
  explored.
\end{itemize}

A similar band-edge shift has been observed earlier in
proton-irradiated BGO-crystals\,\cite{nim206:107} while no such shift
was created by $\gamma$-exposure. A later re-analysis of the data
in\,\cite{nim206:107} revealed that the proton induced damage is
cumulative also in BGO\,\cite{franc-dpf}.

The amplitudes $A_1$ and $A_2$ exhibit some dependence on proton
fluence, which is not perfectly consistent with the above
assumption. It is conceivable, however, that other defect centres,
specific to protons, play some role and have time constants that could
not be resolved from our two universal $\tau_i$ values and the
constant term.
\begin{figure}
\begin{center}
\includegraphics*[height=8.5cm]{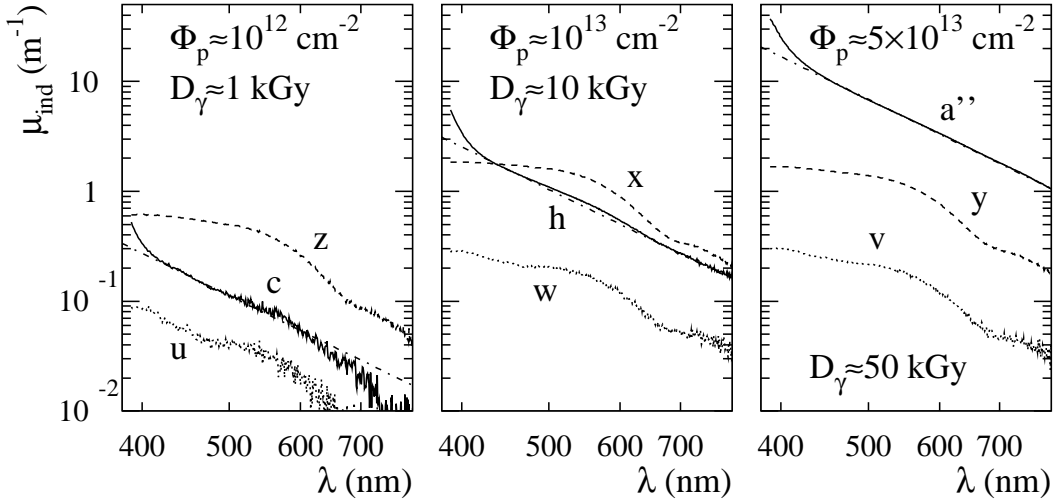}
\end{center}
\caption{Plot of $\muindl$ against $\lambda$ for proton and
  $\gamma$-damaged crystals exposed to different fluences and
  doses. The dot-dashed line shows $\lambda^{-4}$, fitted to the
  proton damage curve. The good agreement is an indication of Rayleigh
  scattering from small centres of severe damage.}
\label{fig17}
\end{figure}

The assumption that the highly ionising fragments, produced by
inelastic hadronic interactions, would be responsible for the observed
specific proton damage implies that there might be very small regions
of severe damage distributed within the crystal. Fig.\,\ref{fig17}
shows $\muindl$, plotted as a function of the wavelength. Above
$\sim$420\,nm the dependency for heavily proton-damaged crystals goes
like $\lambda^{-4}$, which is exactly what would result if the
attenuation was entirely due to Rayleigh-scattering.
In $\gamma$-damaged crystals no such linear dependence appears and
only a structure due to absorption band formation is seen. The
significant crystal-to-crystal variation of $\gamma$-damage is again
evident. The $\muindhosp$ values in Table\,\ref{table1} suggest that
the proton-damaged crystals should rather be compared with the less
damaged $\gamma$-irradiated crystals {\it u}, {\it w} and {\it v}.
For proton damaged crystals, the $\lambda^{-4}$ slope of
Rayleigh-scattering is clearly superimposed on top of the
absorption-band structure from pure ionisation damage. This provides
firm evidence that the specific proton damage is of a very different
nature than what can be caused by purely ionising radiation.  At
fluences of the order of $10^{12}$\,cm$^{-2}$ and below, the
$\lambda^{-4}$ slope starts to get submerged in the absorption band
structure which means that at low fluences it would be difficult to
distinguish the specific proton damage.

The $\muindl$ slope starts to deviate strongly from the $\lambda^{-4}$
behaviour at wavelengths below 420\,nm where the crystals become
strongly absorptive. The Rayleigh scattering formalism is known to
break down when the dimensions of the scattering centres become larger
than $\sim$10\% of the wavelength.  This means that the created defect
centres must have dimensions $\ll$0.4\,${\mu}$m.

The fact that proton damage seems to be connected to very small,
highly damaged, regions strongly points in the direction that the
origin is indeed in the heavily ionising fragments created in
inelastic hadronic interactions.  In this case, the overall dose rate
of the irradiations should not be significant, since the local
instantaneous ionisation of the fragment is overwhelmingly
large. Although the PS beam did not allow to lower the intensity in
order to obtain firm experimental proof for this, all evidence
supports the assumption that the proton damage will be cumulative
irrespective of the dose rate.

\section{Extrapolation to CMS at LHC}

The high-precision ECAL will play a crucial role in the CMS experiment
for the search of the Higgs boson and exploration of possible new
physics beyond the Standard Model.  The calorimeter comprises
$\sim$76000 PbWO$_4$ crystals, which provide hermetic coverage up to a
pseudorapidity $\eta=3.0$. In order to fully profit from the advantage
of a homogeneous calorimeter, namely a very low stochastic term in the
energy resolution, the constant term must be kept low as well. This
can be achieved only by accurate calibration and permanent monitoring
of the crystals. Radiation damage, not compensated by monitoring,
would quickly degrade the calorimeter performance. Therefore, the LT
of each crystal will be calibrated with a 440\,nm laser. In addition
$Z\rightarrow e^+e^-$ and $W\rightarrow e\nu$ events can be used for
{\it in-situ} calibration. However, if the transparency of the
crystals decreases, this will ultimately affect the photostatistics
and thereby increase the stochastic term in an unrecoverable way.

The endcaps (EE) are the parts of the ECAL which are exposed to the
harshest radiation environment. The area where high resolution is
desirable extends to $|\eta|=2.5$.
\begin{figure}[bht]
\begin{center}
\includegraphics*[height=9.5cm]{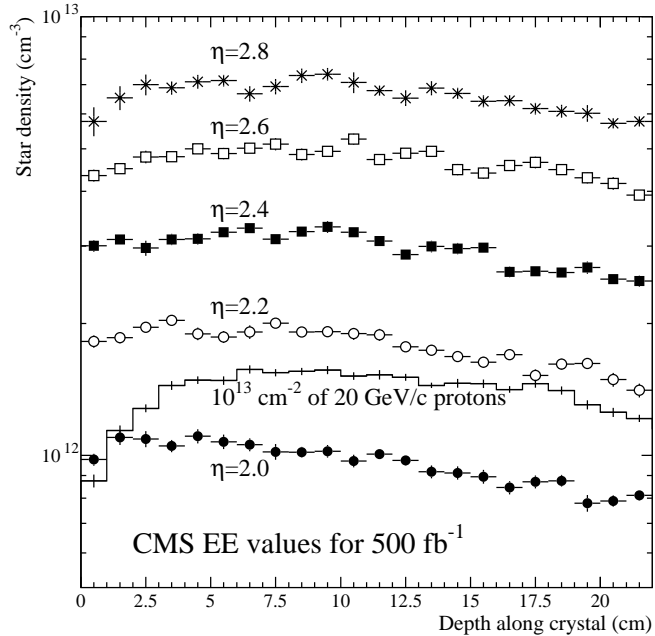}
\end{center}
\caption{Star density at different $\eta$ in the CMS endcap ECAL for
  500\,fb$^{-1}$ integrated luminosity.  The histogram shows the
  average star density in a single crystal irradiated with $10^{13}$
  protons per cm$^2$ in the 20\,GeV/c beam-line.}
\label{fig18}
\end{figure}
The hadron spectrum incident on the EE varies with radius, but it is
always broad and on average much less energetic than the 20\,GeV/c
proton beam used in our tests.  In Fig.\,\ref{fig18} we compare the
simulated star densities at various $\eta$-values in the EE and along
the centre of a crystal exposed to $10^{13}$\,p/cm$^2$ at 20\,GeV/c.
In terms of pure star density, a fluence of
$\Phi_p=10^{13}$\,cm$^{-2}$ of 20\,GeV/c protons would correspond to
the star density expected at $\eta\approx 2.2$ over 10\,years LHC
operation.  However, the characteristics of stars change with
energy. While in our beam test a majority of stars are generated by
high-energy protons, in the EE they are predominantly due to pion and
neutron interactions at energies rarely exceeding 1\,GeV.  At low
energy typically a few nucleons are emitted and one heavy slow recoil
is produced.  With increasing energy, the multiplicity of secondary
particles increases and the mean atomic mass of the heavy nuclear
fragment decreases, while its energy increases slowly.  Energies of
fission fragments are fairly constant but the cross section is
energy-dependent.  These effects can be taken into account only by a
detailed simulation of the hadron-nucleus events followed by transport
of the fragments down to zero energy.  However, prior to such a
simulation, it has to be decided which is the physical process causing
the damage.

We have argued that stars are likely to be at the origin of the
specific proton damage and our experimental results -- especially the
consistency with Rayleigh scattering -- seem to support this. However,
our irradiation data do not tell us by what mechanism the stars cause
the observed damage.
In high-resistivity silicon the hadron damage is known to be due to
the created impurities and, especially, lattice
defects\,\cite{nim491:194}. However, since the natural impurity
concentrations in PbWO$_4$ are much higher, a similar damage mechanism
cannot be assumed without a quantitative consideration.  From
Fig.\,\ref{fig3} we deduce an average of 0.14\,stars/cm$^3$ per unit
of flux. Thus, for $\Phi_p=10^{13}$\,cm$^{-2}$ we obtain an impurity
concentration of about \expfor{3}{-11} per atom.  This is far lower
than the natural density of imperfections in the crystal, which is at
the $10^{-6}$ level.

However, along their path, the recoils collide elastically with atoms
of the lattice and kick these out from their position. Most of the
displacements heal immediately, but some remain stable or metastable.
Simulation methods of fragment transport have been developed for
studies of Single Event Upsets in silicon based electronics
chips\,\cite{nim450:155}. Thereafter they have been successfully
applied to bulk damage studies of silicon
detectors\,\cite{nim491:194}. Silicon, in contrast to PbWO\,$_4$, is a
well known material from this point of view. Thus the adaption of the
simulation to the crystals introduced several uncertainties.  In
particular, the displacement energy, about 20\,eV in silicon, is not
known for PbWO\,$_4$.  While it is likely to be higher, it is unlikely
to be much different from the silicon value.  For our simulations we
assume a value of 50\,eV.  The simulations proceed as described in
\cite{nim491:194,nim450:155} and give on average 2700 displaced atoms
per 20\,GeV/c proton. Counting both the vacancies and the
interstitials as defects, the number doubles. However, at least 90\%
of the vacancy-interstitial pairs will recombine immediately, so a
reasonable assumption for the number of defects created per proton is
$<$500. This gives an upper limit of $\sim10^{-8}$ for the defect
concentration per lattice atom, which still remains about two orders
of magnitude below the natural impurity levels.
\begin{figure}
\begin{center}
\includegraphics*[height=9.5cm]{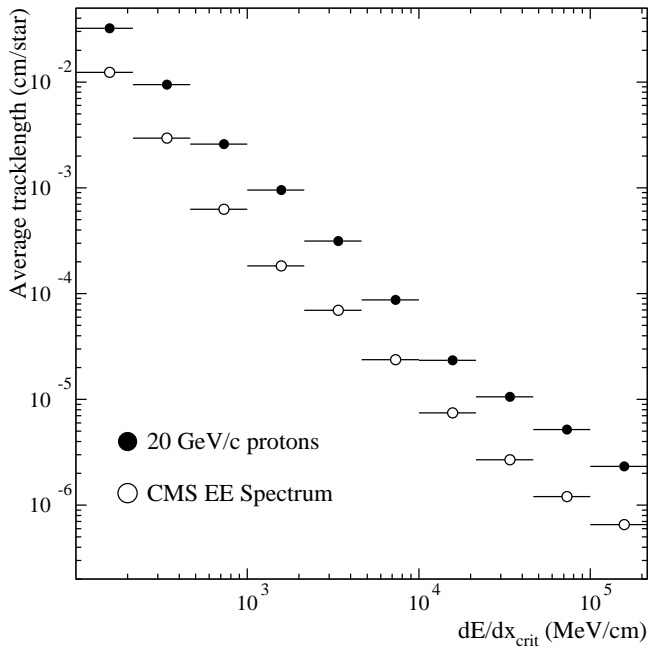}
\end{center}
\caption{Average sum of tracklength per star with ionisation density
  above a given value of (dE/dx)$_{\rm crit}$ for the beam test and
  for CMS endcap ECAL conditions.}
\label{fig19}
\end{figure}

A more plausible damage mechanism seems to be that the created nuclear
fragments cause extremely high local ionisation densities along the
short path which they cover in the crystal.  It can be seen from
Fig.\,\ref{fig1} that a lead ion can reach a maximum $dE/dx$ of
\expfor{5}{5}\,MeV/cm, which is 50000 times larger than for a minimum
ionising particle. It is certain, however, that Pb fragments with
1\,GeV kinetic energy will not be created by hadron-nucleus
collisions.  The most energetic fragments are due to high-energy
induced fission of W and Pb and typically have energies of 100\,MeV or
just below.
Zr and Fe, shown in Fig.\,\ref{fig1}, are good examples of possible
fission products.  Thus the maximum $dE/dx$ of the fragments can be
about \expfor{2}{5}\,MeV/cm.  If these very densely ionising fragments
are responsible for specific hadron damage, then the main question is
what is the critical value ($dE/dx$)$_{\rm crit}$ needed to create a
stable damage centre. If such a value can be identified, then the
amount of damage is quite likely to scale with the total tracklength
of fragments exceeding ($dE/dx$)\,$_{\rm crit}$. This tracklength is
shown in Fig.\,\ref{fig19} for ($dE/dx$)$_{\rm crit}$ values above
100\,MeV/cm. The tracklengths are normalized per star, which allows a
direct comparison of the reactions induced by 20\,GeV/c protons and
those expected in the CMS EE, for which an average spectrum between
$\eta$=1.9--2.8 has been used in the simulation. In order to be
insensitive to the non-uniformity close to the crystal edges, the
tracklengths were recorded only $\pm$5\,mm around the centre line of
the crystal, which corresponds to the region mapped in the LT
measurements.

The difference in tracklength between the 20\,GeV/c proton induced
events and those created by the EE spectrum shows only a weak
dependence on ($dE/dx$)\,$_{\rm crit}$ and amounts on average to a
factor of 4. At ($dE/dx$)\,$_{\rm crit}$ below 1\,GeV/cm, the
tracklength is mostly due to slow protons.  For these it should be
noted that the simulations consider only protons produced by the
stars.  Especially in the CMS EE, but to some extent also in our beam
test, protons produced elsewhere can enter the crystal and produce a
highly ionising track upon stopping\footnote{In the simulations for
  the EE, one crystal within a big matrix was considered, while for
  the beam test a single crystal in infinite void was simulated.}.
For ($dE/dx$)\,$_{\rm crit}>$1\,GeV, heavier fragments dominate
because, as shown in Fig.\,\ref{fig1}, the $dE/dx$ of a proton can
never exceed 1.5\,GeV/cm. It should be pointed out that the results of
Fig.\,\ref{fig19} are not sensitive to our assumption of 50\,eV as the
displacement threshold. That parameter affects only the total number
of produced lattice defects, but not the highly ionising tracks.

With the provision that these simulations are based on an unproven
assumption of the mechanism causing the specific hadron damage,
Fig.\,\ref{fig19} predicts that stars produced by our beam test are on
average a factor of 4 more damaging than those expected in the
EE. With 20\,GeV/c protons the specific hadron damage becomes dominant
at $\sum\Phi_p\approx 10^{13}$\,cm$^{-2}$, when $\muind\approx
2$\,m$^{-1}$. Taking the factor of 4 into account but assuming that
the damage is stable over time, this would correspond to roughly
$\eta$=2.6 after an integrated LHC luminosity of 500\,fb$^{-1}$. If
this extrapolation is correct, the performance of the high-resolution
region of the ECAL will not be compromised by hadron damage during the
first 10\,years of LHC operation.

\section{Conclusions}

We have performed an extensive irradiation campaign, aimed at
verifying the possible existence of specific, hadron-induced damage in
PbWO$_4$ crystals. Emphasis has been put on a careful analysis and
detailed understanding of the irradiation conditions and a long
follow-up of damage recovery.

The very intense radiation environment would have damaged any
front-end electronics or other sensitive equipment, making a
determination of damage on the crystal itself complicated if an {\it
  in situ} determination of LT had been attempted. To avoid such
ambiguities we put our emphasis on a simple procedure aimed solely at
the determination of long-term damage. Due to the high rates which we
were forced to use in order to reach 10-year equivalent LHC fluences,
any determination of short term damage would not have been relevant
for LHC conditions, anyway.

Since no $\gamma$-irradiation data of recent CMS crystals existed at
dose rates corresponding to those caused by the proton beam, we made
comparative irradiations at an intense $^{60}$Co facility.

Only at proton fluences well beyond $10^{12}$\,cm$^{-2}$, i.e. doses much higher than 1\,kGy, we 
start to observe clear differences in the LT-characteristics of crystals exposed to protons and those 
exposed to photons. The main differences are
\begin{enumerate}
\item The proton induced damage increases linearly with fluence while
  the $\gamma$-damage saturates after a few hours of exposure at
  1\,kGy/h.
\item No correlation is seen between the $\muindhosp$, from the
  radiation hardness pre-characterisation with $^{60}$Co photons, and
  the proton induced $\muind$. In the case of our $\gamma$-irradiated
  crystals this correlation is nearly perfect.
\item In proton-irradiated crystals, the band-edge shifts towards
  longer wavelengths and $\muindl\propto \lambda^{-4}$, while the
  band-edge remains stable in the $\gamma$-irradiated crystals and a
  clear absorption-band structure appears in $\muindl$.
\item After $\sum\Phi_p$=\expfor{5.4}{13}\,cm$^{-2}$,
  i.e. $\sim$100\,kGy, we observe $\muind\approx$15\,m$^{-1}$ and a
  linear rise with fluence. For a crystal with similar
  radiation-hardness characteristics, we observe after a $\gamma$-dose
  of $\sim$50\,kGy only $\muind$=0.3\,m$^{-1}$ and no indication of
  further increase towards higher doses.
\end{enumerate}

It was not possible to raise the $\gamma$-intensity beyond 1\,kGy/h,
in order to obtain a comparison with our highest proton beam
intensities. It is usually claimed that the saturation level of
$\gamma$-damage increases with dose rate. For protons, however, we
observe no difference between rates of $10^{12}$\,cm$^{-2}$h$^{-1}$
and $10^{13}$\,cm$^{-2}$h$^{-1}$.

A thorough exploration of fitting various functions to the recovery
data showed, that Eq.\,\ref{timefit} is suitable to fit all data
accumulated so far. In particular we discovered that the quality of
the fits does not suffer, if the time constants, $\tau_i$, are
universal with values of \tauyks\,days and \taukaks\,days.  Only the
absolute amount of damage is very different for the two radiation
types, which is reflected in a different dose dependence for the
fitted amplitudes $A_i$, as shown in Fig.\,\ref{fig13}.

Our results provide strong evidence that proton exposure of PbWO$_4$
crystals causes damage that cannot be reproduced with
$\gamma$-irradiation. This damage seems to be independent of rate and
increases linearly with accumulated fluence. We strongly emphasise
that the difference we observe starts to manifest itself only at
proton fluences well beyond $10^{12}$\,cm$^{-2}$.  Below this fluence
the ``normal'' $\gamma$-like damage, due to minimum ionising tracks,
is dominating to such an extent that it is probably impossible to
experimentally see any difference between protons and photons.

The observation, that $\muindl$ of proton damaged crystals is
consistent with Rayleigh scattering from objects much smaller than
0.4\,$\mu$m, supports our assumption that the specific proton damage
is due to the very dense local ionisation of fragments created in
inelastic nuclear collisions. In this case it is likely that the
20\,GeV/c proton beam used in our tests is somewhat too pessimistic
with respect to the hadron spectrum expected in the CMS ECAL. Our
simulations indicate that the damage ratio between 20\,GeV/c protons
and the CMS ECAL hadron spectrum, per unit star density, might be
about a factor of 4. But this remains to be verified by further
irradiations with low-energy pions, which provide a better
representation of the ECAL conditions.

\appendix

\section{Induced radioactivity}
\label{app-indact}

The creation of nuclear fragments during hadron irradiation causes the
crystals to become radioactive. Since PbWO$_4$ includes the heavy
elements W and Pb, the 20\,GeV/c proton beam is able to produce a huge
number of different nuclides.  Most of these are either stable or
short-lived, but a fraction of them have long half-life and are
responsible for the remnant radioactivity of the irradiated
crystals. According to simulations, there are about 1800 different
nuclide species with an activity above 1\,Bq/cm$^3$ at the end of an
exposure to $10^{13}$\,p/cm$^2$. Even after 50\,days of cooling, 150
nuclides remain above this activity threshold.

The predictive simulation of crystal radioactivation proceeded in
three steps. First, during the full FLUKA simulation described before,
all residual nuclides produced in the crystal were recorded. This
stage relies on the accuracy of the FLUKA models to describe the
hadron-nucleus interaction in full detail. Earlier dedicated work has
shown, that an accuracy better than a factor of 2 can be expected for
heavy materials\,\cite{marc5,marc6}.  In a second stage the evolution
over time of the produced nuclide inventory was calculated with the
DeTra code\,\cite{detra}. This calculation is analytical and thus
exact within the limits of accuracy given by the available nuclear
data files.  DeTra produces a table of all nuclide activities as a
function of cooling time.  For the final phase of the simulation these
activities were converted into an emitted $\gamma$-spectrum which was
then used to calculate $\iadr$ at our measurement point by integrating
over the entire crystal and weighting with the simulated star density
within the crystal. The integration was performed via a detailed
photon transport with FLUKA.

\begin{figure}
\begin{center}
\includegraphics*[height=13cm]{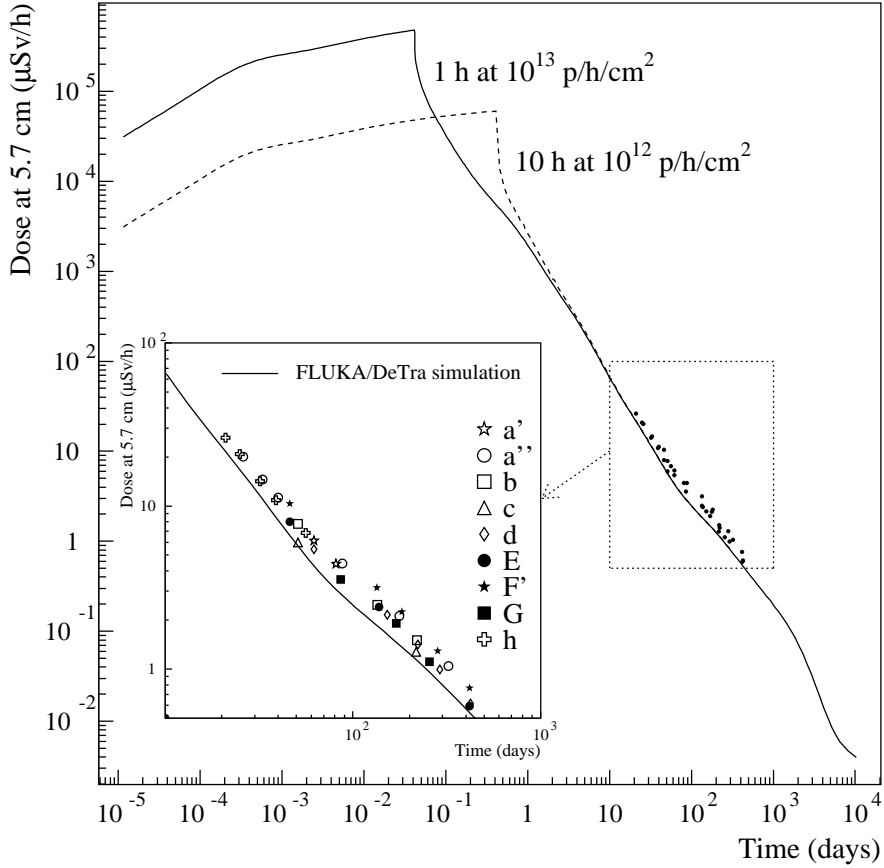}
\end{center}
\caption{Comparison of simulated and measured radioactivity as a
  function of cooling time. All data have been renormalised to
  $\sum\Phi_p$=$10^{13}$\,cm$^{-2}$h$^{-1}$ using the fluence
  information from Al foils, as given in the $\sum\Phi_p$ column of
  Table\,\protect\ref{table3}.}
\label{fig20}
\end{figure}

Figure\,\ref{fig20} compares simulated $\iadr$ and all our
measurements. For this purpose all measurements have been rescaled to
$\sum\Phi_p=10^{13}$\,cm$^{-2}$, using the total proton fluence, as
determined from activation of aluminium foils. The two curves in
Fig.\,\ref{fig20}, corresponding to different beam intensities but the
same integral $\sum\Phi_p$, show that by the time our measurements
could begin only the value of $\sum\Phi_p$ matters, not the rate.
Thus a rescaling by $\sum\Phi_p$ is justified for our comparison.

In the small insert of Fig.\,\ref{fig20} the doses from individual
crystals are compared with the calculated $\iadr$. We observe a
systematic underestimate in the simulation.  On average the deviation
amounts to 29\% and even at worst the disagreement remains below a
factor of 2, which still can be considered a very good agreement for
such a prediction. We also have to recall two uncertainties in the
measured values with respect to the simulation, namely the unknown
efficiency of the Automess 6150AD6 above 1.3\,MeV, which we probably
underestimate, and a slight uncertainty in the size and position of
the sensitive element within the device.

\begin{table}
\begin{center}
\begin{tabular}{c|c|c|c} 
                       & Front  & Middle  & Rear \\ \hline
Measured               & 7.67   & 10.4    & 8.64 \\
Calculated             & 5.09   &  6.58   & 5.48 \\ \hline
1.58$\times$Calc/Meas  & 1.05   & 1.00    & 1.00 \\ \hline
\end{tabular}
\end{center}
\caption{Measured and calculated $\iadr$ ($\mu$Sv/h) close to the
  front, in the middle and close to the rear of crystal {\it a"} (see
  Table\,\protect\ref{table3}). The values correspond to 177\,days
  cooling after the last irradiation.}
\label{table5}
\end{table}

We used the $\iadr$ measurements also to cross-check the longitudinal
star density simulations shown in Fig.\,\ref{fig3}. For this purpose
we measured $\iadr$, in addition to the middle, also close to the
front and back of crystal {\it a"}, which had been exposed to
$\sum\Phi_p$=\expfor{5.4}{13}\,cm$^{-2}$.  The results are given in
Table\,\ref{table5}. Correcting for the systematic underestimate of
the simulations in the case of {\it a"}, the relative
front--middle--back ratios are very well reproduced\footnote{The
  values in front and back are particularly sensitive to the exact
  longitudinal position and the assumed size of the sensitive element
  in the 6150AD6.}.  This good agreement increased our confidence that
the simulations, summarised in Figs.\,\ref{fig3} and \ref{fig4},
describe well the conditions within the crystal.

\section{Technical aspects of $\Phi_p$ determination}
\label{app-pflux}

\begin{figure}
\begin{center}
\includegraphics*[height=10cm]{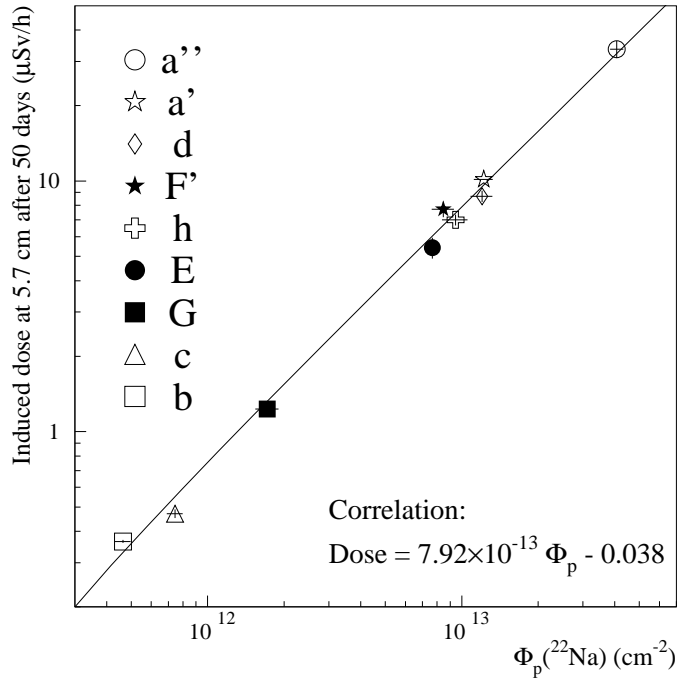}
\end{center}
\caption{Correlation of measured $\iadr$ of the crystals and $\Phi_p$
  as determined from $^{22}$Na activity in the Al foils.}
\label{fig21}
\end{figure}
The most reliable standard way to determine the fluence of the
incident proton beam ($\Phi_p$) is by means of aluminium activation
measuring the production of $^{24}$Na.  This is the isotope for which
cross sections are known most reliably.  However, the cross section
Al(n,$\alpha$)$^{24}$Na above $\sim$6\,MeV is much larger\footnote{The
  Al(n,$\alpha$)$^{24}$Na cross section exceeds 100\,mb between
  10--30\,MeV and then decreases towards higher
  energies\,\protect\cite{in2001:12}.} than the Al(p,X)$^{24}$Na cross
section at 20\,GeV (8.7\,mb). Since the crystal produces a large flux
of neutrons above this threshold energy, $^{24}$Na is not a useful
isotope to monitor $\Phi_p$, on top of the fact that its 15\,h
half-life makes it impractical for our purpose.  A more suitable
isotope is $^{22}$Na.  Due to its long half-life it integrates the
fluence, and is insensitive to short-term beam fluctuations.

The cross section for Al(p,X)$^{22}$Na at $\sim$20\,GeV is not known
as accurately as for Al(p,X)$^{24}$Na.
Data in\,\cite{landbor} exhibit a significant spread but allow us to
estimate 11.2$\pm$1.6\,mb. The ratio $^{24}$Na/$^{22}$Na in a foil,
irradiated in the absence of a crystal, gave 10.4\,mb. Since this is
based on the much better known production cross section of $^{24}$Na,
we use 10.4\,mb for the fluence analysis.  However, we have to point
out that even without a crystal, backscattered radiation in IRRAD1
increases the production of $^{24}$Na more than of $^{22}$Na. Thus the
10.4\,mb is likely to represent a lower bound. Consequently, $\Phi_p$
is likely to be rather over- than underestimated.

While certainly smaller than for Al(n,$\alpha$)$^{24}$Na, the cross
sections for Al(n,X)$^{22}$Na are not known well enough to allow for
an accurate analysis of the neutron contribution. With our best
knowledge of the cross sections we obtain from the FLUKA simulation a
11\% contribution to the $^{22}$Na activity from non-beam
particles. This we propagate into our estimate of the incident proton
fluence.

The aluminium foils were cut to a size of $24\times 24$\,mm$^{2}$,
which roughly corresponds to the average crystal dimensions. Thus the
foils were averaging the whole fluence incident on the crystal. In a
single crystal, however, protons close to an edge contribute on
average less star density, thus damage and activation, because a
larger fraction of the hadronic cascade escapes, when compared to
centrally impinging ones.  In fact, the leakage of the cascade causes
that the radiation intensity drops towards the edges even with a
uniform beam, for which the star density in the centre is about 17\%
higher than the average over the whole crystal. However, would the
crystal be surrounded by other crystals, also covered by the beam,
then these would compensate for the leakage and the profile would be
flat and the total star density higher. Thus the relation between
$\Phi_p$, determined with the aluminium foil, and the star density
depends -- in addition to the beam characteristics -- also on the size
and surroundings of the crystal.

The activation of the crystal by inelastic nuclear interactions is
roughly proportional to the star density inside the
crystal. Unfortunately, this does not provide an absolute
determination of $\Phi_p$, because star density and $\iadr$ can be
related to each other only through simulations, which introduce too
large an uncertainty, as discussed in detail in
appendix\,\ref{app-indact}.
However, since for all our irradiations the conditions within the
crystals are the same, the effects of simulations cancel out if we
consider only the relative differences between irradiations. Thus we
expect a near to perfect correlation between $\iadr$ and the $^{22}$Na
activity of the aluminium foils. The only deviations could arise from
significant non-uniformities of the beam spot or from a severe
misalignment of the crystal during irradiation and, to a lesser
extent, from the decay corrections, for which recourse to simulation
is unavoidable.  This correlation between $\Phi_p$ and $\iadr$, shown
in Fig.\,\ref{fig21}, provided a very valuable cross-check of our
fluence measurements. The $\iadr$ measurements were taken at different
times, but all values have been corrected to 50\,days after
irradiation using the simulated decay of radioactivity in the
crystals. As expected, the correlation is almost perfect, indicating
that the activation foils give reliable and consistent fluence
values. There are no signs of severe anomalies in any individual
irradiation.

Due to a faulty magnet in the PS, the early irradiations took place
with a 20\,GeV/c proton beam while later exposures (crystals {\it a"}
and {\it h}) were performed at 24\,GeV/c.  The energy dependence of
the Al(p,X)Na$^{22}$ cross section is negligible in this
regime. However, as shown in Fig.\,\ref{fig3}, the 20\% increase in
proton momentum leads to more intense showering and thus to a higher
radiation field within the crystal. The effect is very similar for all
hadronic radiation quantities plotted in Fig.\,\ref{fig3} but we use
the 8.6\% increase in star density, averaged over the crystal
length. The LT measurements also average along the whole crystal and
thus justify a simple correction by simply rescaling $\Phi_p$, as
determined from the aluminium foils, by a factor of 1.086 for the two
crystals irradiated with 24\,GeV/c protons.  In the case of ionising
dose the effect is larger because the electromagnetic cascade is much
better contained even in a single crystal.

\section{Influence of beam non-uniformities}
\label{app-nonunif}

\begin{figure}[bht]
\begin{center}
\includegraphics*[height=9.5cm]{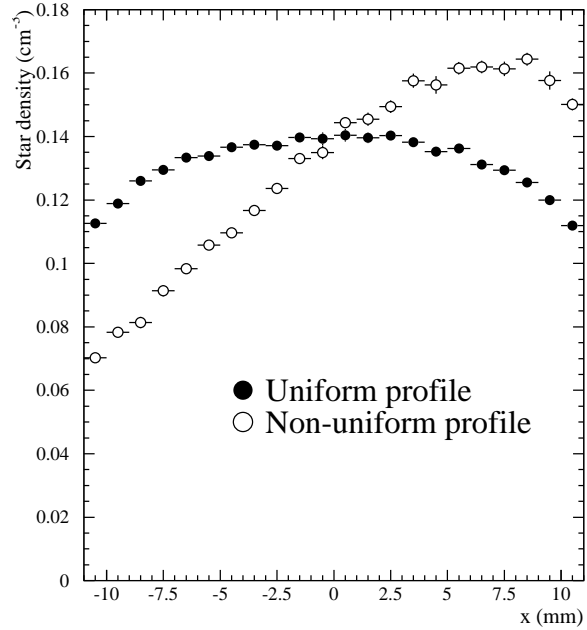}
\end{center}
\caption{Transverse star-density profiles across a crystal for a
  perfectly uniform beam and for a beam with 5 times lower intensity
  at $x$=-12\,mm than at $x$=12\,mm.}
\label{fig22}
\end{figure}
In order to study the effect of non-uniformities of the beam, we
assumed that the beam intensity drops linearly from one side of the
crystal to the other\footnote{A determination of the beam profile
  using OSL (Optically Stimulated Luminescence) films showed that the
  real non-uniformities are not linear but exhibit a bump-like maximum
  on one side.}. Crystal {\it E} was identified as having received the
most non-uniform exposure.  We adjusted the slope such that the
$\iadr$ values measured on {\it E} were roughly reproduced and ended
up with a factor of 5 intensity difference across the crystal front
face.  A full FLUKA simulation was done with such a severe
non-uniformity.  Fig.\,\ref{fig22} compares the lateral profiles of
star density obtained from this simulation and the normal uniform
one. Although a larger fraction of beam hits close to the edge, the
total number of stars differs only by a few percent.  Due to
cascading, the variation in star density is much smaller than the
factor of 5 used as incident beam intensity. We also observe that in
the centre of the crystal the effects of non-uniformity are at a
minimum, which suggests that, provided the crystal alignment is the
same in all LT measurements, even for the most severe beam
non-uniformity encountered the influence on the results remains
negligible in comparison to other uncertainties.
\begin{figure}
\begin{center}
\includegraphics*[height=9.5cm,bb=0 50 567 567]{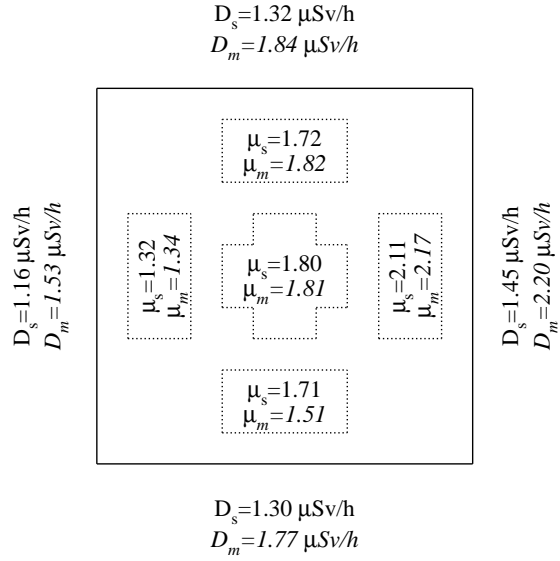}
\end{center}
\caption{Effect of beam non-uniformity in case of crystal {\it E}.
  Compared are the $\iadr$ at 4.5\,cm distance from each long side and
  the $\muind$-values in the centre and close to each side.  The
  measured values are labeled with '$m$' and the simulated ones with
  '$s$'.  }
\label{fig23}
\end{figure}

The factor of 5 beam intensity difference used in the simulations of a
non-uniform beam, as described before, was adjusted so that the
relative\footnote{The absolute value of $\sum\Phi_p$ is always derived
  from the aluminium foils.} variation of $\iadr$ observed on the most
non-uniform crystal, {\it E}, was reproduced when the simulated data
were used to calculate $\iadr$ on each side.

The same simulated data were also used to integrate the star density
along the light paths of the LT measurements. Assuming that stars
cause the damage, the fit of Fig.\,\ref{fig15} allows to translate
these star densities into $\muind$ values, which can be compared with
the measured values on absolute scale.

Figure\,\ref{fig23} shows such comparisons for $\iadr$ and $\muind$ in
the case of crystal {\it E}. The systematic underestimate of $\iadr$
is consistent with Fig.\,\ref{fig20}, where the simulated
$\iadr$-values for crystal {\it E} are also slightly lower than the
measurements.  However, it must be emphasised that for such a
simulation an agreement better than a factor of two, is remarkable.

The simulated $\muind$ values are in perfect agreement with
measurements, except for a slight up-down asymmetry. This is most
probably related to a beam non-uniformity in the vertical direction,
while the simulated beam intensity was varied only horizontally.

In relative terms, the non-uniformities in both $\muind$ and $\iadr$
are very well reproduced by just considering the spatial variation of
star density within the crystal.

\section*{Acknowledgements}
The proton irradiations, performed at CERN, were made possible only by
the efforts of R. Steerenberg, who provided us with the required PS
beam conditions. The help of M. Glaser and F. Ravotti in operating the
irradiation and dosimetry facilities is gratefully acknowledged. The
support and practical help of S. Baccaro and A. Cecilia during the
$\gamma$-irradiations at ENEA-Casaccia was essential and is gratefully
acknowledged, as is the contribution of E.\,Auffray who provided us
with the crystals after having characterised their $\gamma$-radiation
hardness at the Geneva Hospital.

\end{document}